\documentclass[twoside,leqno,twocolumn]{article}

\usepackage[letterpaper]{geometry}

\usepackage{ltexpprt}
\usepackage{url}
\usepackage{authblk}
\usepackage{booktabs}
\usepackage{amsmath, amssymb}
\usepackage{mathtools}
\usepackage{breqn}

\newtheorem{definition}{Definition}

\newtheorem{observation}{Observation}
\usepackage{braket,amsfonts,amsopn}
\usepackage{algorithm}
\usepackage{algorithmic}
\usepackage{xfrac}
\usepackage{color}

\DeclareMathOperator{\LN}{L_{\mathbb{N}}}
\DeclareMathOperator{\BE}{BE}
\DeclareMathOperator{\KL}{KL}
\usepackage{graphicx}
\usepackage{subcaption}

\newcommand{\method}{DPGS }
\newcommand{\kGs}{k-Gs }

\begin{document}

\title{\Large Summarizing graphs using the configuration model}

\author[1]{Houquan Zhou\thanks{Email: zhouhouquan18@mails.ucas.ac.cn}}
\author[1]{Shenghua Liu}
\author[2]{Kyuhan Lee}
\author[2]{Kijung Shin}
\author[1]{Huawei Shen}
\author[1]{Xueqi Cheng}

\affil[1]{Institute of Computing Technology, Chinese Academy of Sciences}
\affil[2]{Graduate School of AI, KAIST}
\renewcommand\Authands{ and }

\date{}

\maketitle


\fancyfoot[R]{\scriptsize{Copyright \textcopyright\ 2021 by SIAM\\
Unauthorized reproduction of this article is prohibited}}





\begin{abstract} {\small
Given a large graph, how can we summarize it with fewer nodes and edges while maintaining its key properties, e.g. node degrees and graph spectrum?
As a solution, graph summarization,
which aims to find the compact representation
for optimally describing and reconstructing a given graph,
has received much attention, and numerous methods have been developed for it.
However, many existing works adopt \textit{uniform reconstruction} scheme, which is an unrealistic assumption as most real-world graphs have highly skewed
node degrees, even in a close community.
Therefore we propose a degree-preserving graph summarization model, DPGS, with a novel reconstruction scheme based on configuration model. By optimizing the Minimum Description Length of our model, a linearly scalable algorithm is designed with hashing technique.
We theoretically show that the minimized reconstruction error of the summary graph can bound the perturbation of graph spectral information.
Extensive experiments on real-world datasets show that \method yields more accurate summary graphs than the well-known baselines. Moreover, our reduced
summary graphs can effectively train graph neural networks (GNNs) to save computational cost.

}\end{abstract}

\section{Introduction} 
Recent years have witnessed the explosive growth of data size, and large-scale graphs have become ubiquitous, 
including social networks, computer networks,  and protein interactions network.
Since they are hard to process, analyze, and understand, this poses significant challenges to graph mining applications.

An effective technique to tackle such challenges is \textbf{graph summarization}.
Given a graph $G$, it aims to find a compact representation of $G$ in the form of a  \textit{summary graph} with \textit{supernodes} (i.e., subsets of nodes in $G$) and \textit{superedges} (i.e., subsets of edges in $G$).


Generally, graphs are expected to be reconstructed from summary graphs
by a summarization model, 
and thus the reconstruction scheme is the heart of most summarization models.
However, closely related works \cite{lefevre2010grass, riondato2017graph, beg2018scalable, lee2020ssumm} reconstructed connections of original nodes in a uniform assignment of superedges, namely \textit{uniform reconstruction} scheme. 
And many real-world graphs always contain proper communities of degree-skewed nodes~\cite{araujo2014beyond}, that is, ~popular nodes or leaders 
in a community usually have more connections.
Thus the summarization model based on uniform reconstruction scheme always results in
a large margin between the reconstructed graph and the original graph in terms of node degrees, 
losing related properties such as graph spectrum (i.e.~eigenvalues),  and nodes' authorities and hubness~\cite{kleinberg1998authoritative}.

Therefore, we propose a \textbf{D}egree-\textbf{P}reserving \textbf{G}raph \textbf{S}ummarization model, named \method, which assigns superedges proportional to 
node degrees known as configuration model. 
Such a configuration-based assignment as a scheme (CR scheme) can generally be plugged in and improve existing related summarization methods.
\method uses minimum description length (MDL) from 
information theory as a principle to minimize
the cost of summary graph and reconstruction error.
Theoretically, we show that \method can bound the perturbation of Laplacian's eigenvalues by minimized reconstruction error.
A fast algorithm is designed for \method to summarize large graphs
based on LSH (Locality Sensitive Hashing) to group candidate nodes, and perform greedy merging within groups. 

Empirical study on synthetic datasets (random graphs with both uniform and skewed degree distributions) validates that the proposed scheme can reconstruct the original graph
more accurately than uniform reconstruction scheme, especially 
in highly skewed graphs. 
Extensive experiments on 8 real-world datasets show that \method algorithm outperforms the state-of-the-art algorithms. 
Furthermore, we also show that our summary graphs can efficiently yet effectively help to train a graph neural network, while preserving high accuracy in node classification task. 

In summary, our contributions includes:
\begin{itemize}
    \item \textbf{Novel reconstruction scheme}: We propose a graph summarization model, named \method, with a novel reconstruction scheme based on configuration model. 
    We theoretically show that our \method can bound the perturbation of graph spectrum with the reconstruction error. 
    
    \item \textbf{Compatibility}: Our pro scheme can be generally applied 
    to other graph summarization models, and 
    improve the summary quality.
    
    \item \textbf{Effectiveness}: 
    Experiments on both synthetic and real-world graphs verify the superiority of the proposed reconstruction scheme, and show
    that \method algorithm outperforms the state-of-the-art methods with
    better summary. 
    Moreover, summary graphs can help to train graph neural networks
    efficiently yet effectively.
    
    
    \item \textbf{Scalability}: 
    Our \method algorithm runs fast, and theoretical analysis shows that the complexity is linear in number of edges.
\end{itemize}

\paragraph{Reproducibility}: Our \method algorithm is open-sourced at \url{https://github.com/HQJo/DPGS}.

\begin{figure*}[t]
    \vspace{-4mm}
    \begin{subfigure}{0.3\textwidth}
        \centering
        {\includegraphics[width=\linewidth]{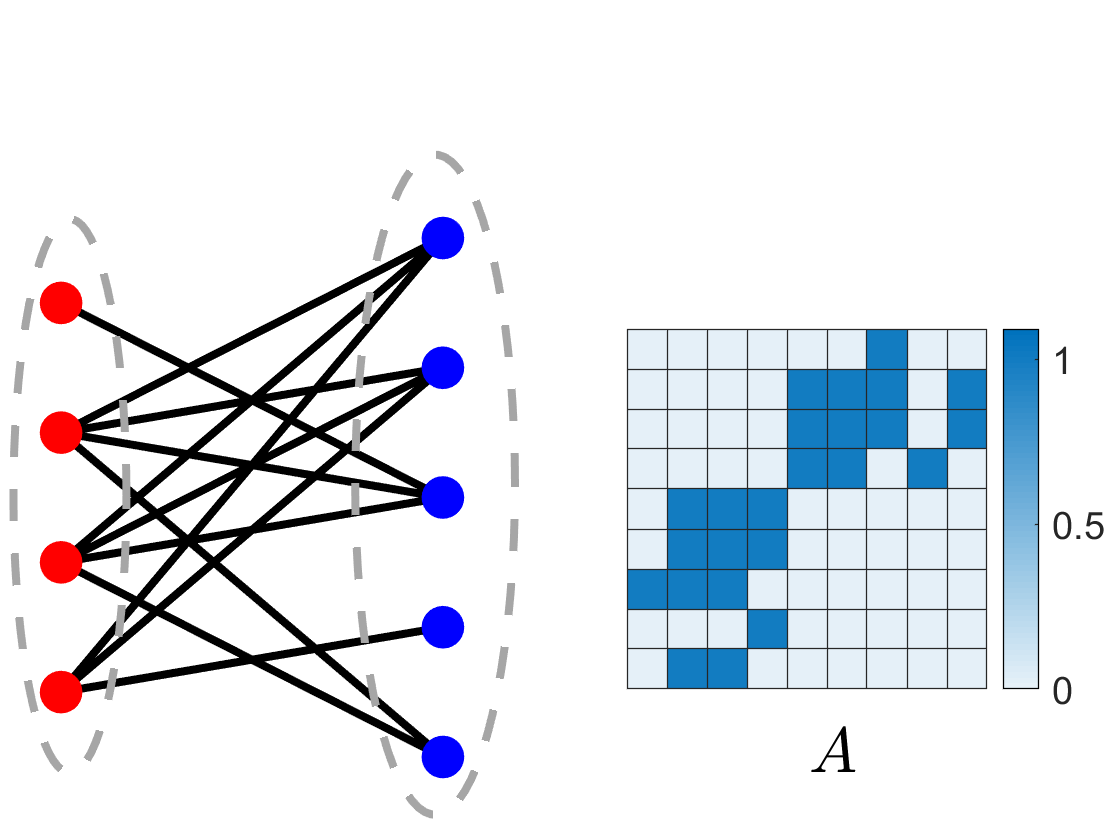}}
        \caption{Original graph}\label{fig:recon_orig}
    \end{subfigure}
    \quad
    \begin{subfigure}{0.3\textwidth}
        \centering
        {\includegraphics[width=\linewidth]{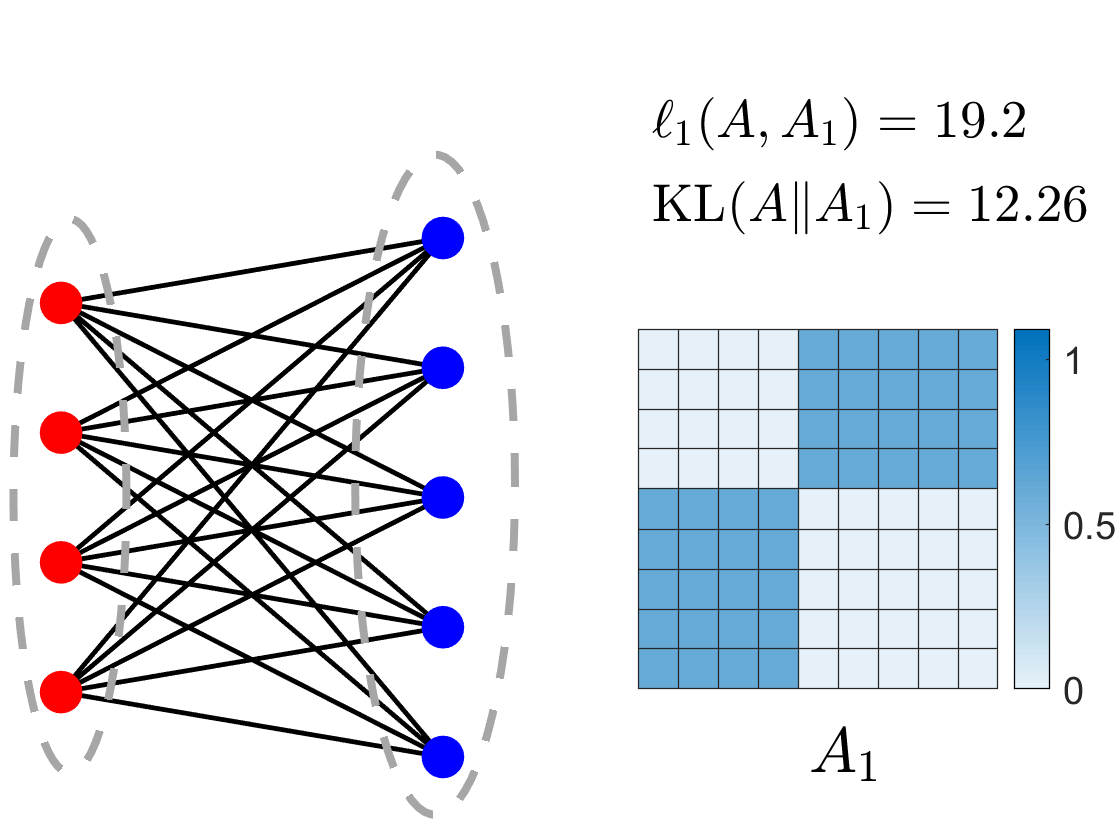}}
        \caption{Reconstructed (uniform)}\label{fig:recon_uniform}
    \end{subfigure}
    \quad
    \begin{subfigure}{0.3\textwidth}
        \centering
        {\includegraphics[width=\linewidth]{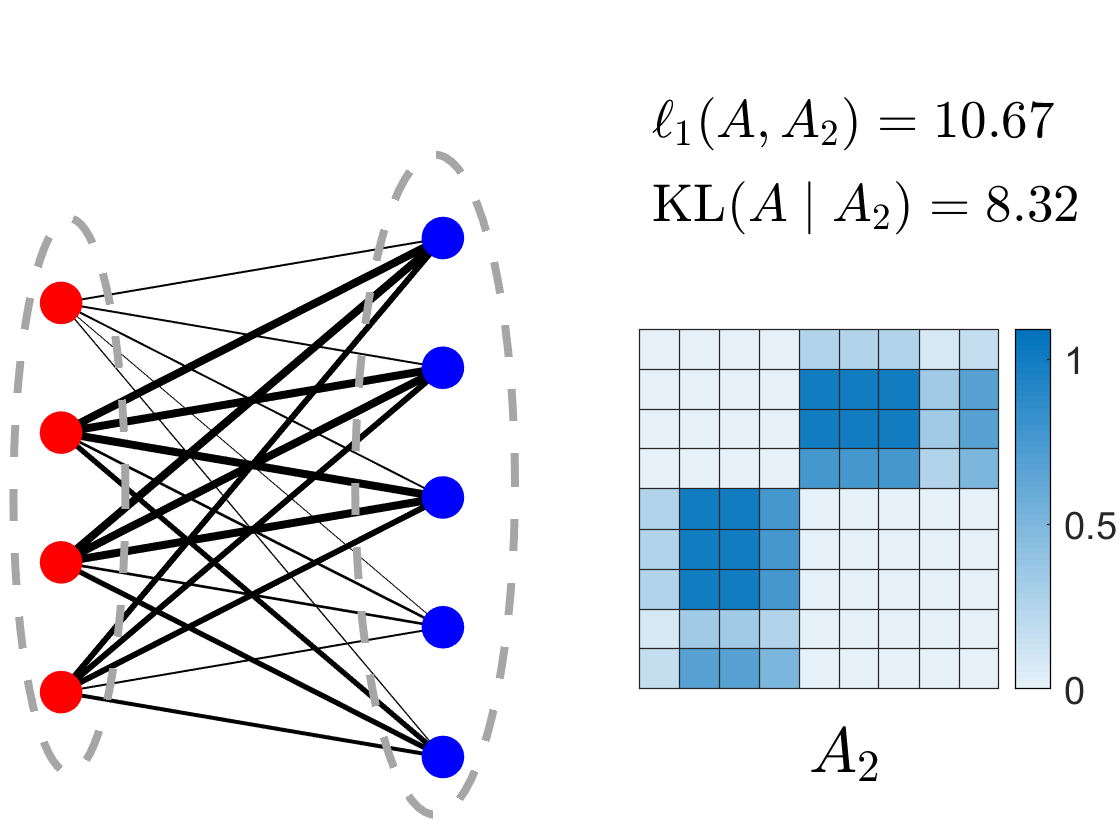}}
        \caption{Reconstructed (CR)}\label{fig:recon_config}
    \end{subfigure}
    \caption{Comparison of two reconstruction schemes. The original graph and reconstructed graphs using two reconstruction schemes are shown in the figure. Heatmaps of the corresponding adjacency matrices are plotted right to graphs. Red nodes and blues nodes are two supernodes and the width of edges reflects the edge weight. From both visualization and quantitative metrics perspective, the proposed CR scheme achieves better reconstruction result. 
    }\label{fig:toy_example}
\end{figure*}
\section{Related Work}
Graph summarization methods can be categorized into many categories, such as grouping based, compression based and influence based, etc. See this excellent survey~\cite{liu2018graph} for more knowledge about this topic. In this section, we categorize some graph summarization problems according to their objective functions.

\paragraph{Error of adjacency matrix as objective function}: This kind of methods try to minimize some error metrics between the original and reconstructed adjacency matrices.
\kGs~\cite{lefevre2010grass} aimed to find a summary graph with at most $k$ supernodes, such that the L1 reconstruction error is minimized.
Riondato~\cite{riondato2017graph} revealed the connection between geometric clustering problem and graph summarization problem under multiple error metrics (including L1 error, L2 error and cut-norm error), and proposed a polynomial-time approximate graph summarization method based on geometric clustering algorithms.
Beg~\cite{beg2018scalable} developed an randomized algorithm SAA-Gs using weighted sampling and count-min sketch~\cite{cormode2005improved} techniques to find promising node pairs efficiently.

\paragraph{Total edge number as objective function}: In this kind of methods, the objective function is defined as number of edges in summary graph plus edge corrections.
In~\cite{navlakha2008graph}, Navlakha proposed two algorithms: \textsc{Greedy} and \textsc{Randomized}. The former considers all possible node pairs at each step, and merges the best pair $(u, v)$ which results in the greatest decrease of the total edge number. The latter samples a supernode as $u$ randomly at each step, checks all other supernodes, finds the best $v$ and merges them together. This process continues until the summary graph is smaller than a given size. However, both algorithms are computationally expensive.
To address this problem, SWeG~\cite{shin2019sweg} reduces the search space by grouping supernodes according to their shingle values firstly, and only considers merging node pairs in same groups. Combined with parallelization technique, SWeG are able to scale to large graphs up to ten billions of edges.

\paragraph{Encoding length as objective function}: This kinds of methods often adopt MDL principle and use total encoding length as objective functions.
The typical paradigm is to propose an encoding scheme and optimize the total description length under the scheme. 
LeFevre~\cite{lefevre2010grass} formulated the graph summarization problem \textsc{Gs} using MDL principle, and proposed three algorithms, \textsc{Greedy}, \textsc{SamplePairs} and \textsc{LinearCheck}. However, the encoding scheme of Gs is highly redundant and the time complexity is high.
Lee~\cite{lee2020ssumm} designed a dual-encoding scheme and proposed a sparse summarization algorithm SSumM which reduces the number of node and sparsifies the graph simultaneously. By dropping sparse edges and encoding them as errors, SSumM are able to obtain a compact and sparse summary graph.
Different from methods mentioned above, VoG~\cite{koutra2014vog} adopted an vocabulary-based encoding scheme, which encodes the graph using frequent patterns in real-world graphs, such as clique, star and bipartite core, etc.

Methods mentioned above mainly focus on static simple graphs. There are works aiming to summarize other types of graphs, including dynamic graphs~\cite{shah2015timecrunch, adhikari2017condensing, qu2014interestingness}, attributed graphs~\cite{khan2017set, hassanlou2013probabilistic, wu2014graph} and streaming graphs~\cite{tang2016graph, ko2020incremental}, etc.

\section{Proposed Method}
In this section, we will describe our DPGS model. To enhance the readability, we list frequently used symbols in Table~\ref{tab:notations}.
\begin{table}[t]
    \caption{Notations}
    \label{tab:notations}
    \begin{center}
        \begin{tabular}{p{0.2\columnwidth}p{0.6\columnwidth}}
            \toprule
            Notation & Description \\
            \midrule
            $G$   & Input simple undirected graph \\
            $\mathcal{V}, \mathcal{E}$ & Node set and edge set of $G$ \\
            $n, m$ & Size of $\mathcal{V}$ and $\mathcal{E}$ \\
            $A_{n\times n}$ & Adjacency matrix of $G$ \\
            $G_S$ & Summary graph \\
            $\mathcal{V}_S, \mathcal{E}_S$ & Supernode set and superedge set of $G_S$ \\
            $n_s, m_s$ & Size of $\mathcal{V}_S$ and $\mathcal{E}_S$ \\
            $(A_S)_{n_s\times n_s}$ & Adjacency matrix of $G_S$ \\
            $A_{n\times n}^\prime$ & Reconstructed adjacency matrix \\
            $d_i$ & Degree of node $i$ \\
            $D_k$ & Degree of supernode $S_k$, i.e., the sum of degrees of nodes in $S_k$ \\
            \bottomrule
        \end{tabular}
    \end{center}
\end{table}

\subsection{Reconstruction Scheme}
Given a summary graph, we can reconstruct the graph by a graph summarization model.
Existing summarization model reconstructs the adjacency matrix using \textit{uniform reconstruction} scheme~\cite{lefevre2010grass}. 
\begin{definition}[Uniform Reconstruction Scheme]\label{dfn:unif_recon}
Denote $A_S$ and $A^\prime$ as the adjacency matrix of the summary graph and the reconstructed graph respectively. The uniform reconstruction scheme calculates $A^\prime(i, j)$ as follows:
\begin{equation}\label{equ:expected_recon}
    A^\prime(i, j) = \begin{dcases}
        \frac{A_S(k, l)}{|S_k|\cdot |S_l|}\, & k \ne l \\
        \frac{A_S(k, l)}{|S_k|(|S_k|-1)},\, & k = l \, .
    \end{dcases}
\end{equation}
where $S_k$ and $S_l$ are supernodes to which node $i$ and node $j$ belong, respectively.
\end{definition}

It can be seen from Equation~\eqref{equ:expected_recon} that the edges between two supernodes $S_k$ and $S_l$, i.e.~$A_S(k, l)$, are equally assigned to each node pair between them, and each node pair has the same connection weight. 
Thus, this approach corresponds to $G(n, m)$ random graph model (or Erd\H {o}s-R\'enyi model equivalently)~\cite{erdos59a}. 
However, the real-world graphs have a \textbf{skewed} nature. Thus, this uniform reconstruction scheme is not suitable for real-world graphs.

Different to the uniform-reconstruction scheme, we reconstruct $A^\prime$ based on degrees of nodes:
\begin{definition}\textsc{(CR Scheme)}\label{dfn:conf_recon}
Denote $A_S$ and $A^\prime$ as the adjacency matrix of the summary graph and the reconstructed graph respectively. The configuration-based reconstruction scheme (CR scheme) calculates $A^\prime(i, j)$ as follows:
\begin{equation}
    A^\prime(i, j) = \frac{d_i}{D_k} A_S(k, l) \frac{d_j}{D_l} \,. 
\end{equation}
where $S_k$ and $S_l$ are supernodes to which node $i$ and node $j$ belong respectively. $d_i$ and $D_k$ denote the degree of node $i$ and supernode $S_k$, and analogue for $d_j$ and $D_l$.
\end{definition}

In this way, the reconstructed edge weight $A^\prime(i, j)$ is proportional to the product of endpoints' degrees. This approach corresponds to the configuration model~\cite{newman_networks_2010}, which have proved successful in modularity-based community detection~\cite{newman_modularity_2006}.

Note that the proposed CR scheme is able to preserve the degrees.
\begin{property}[Degree Preservation]
    \begin{equation}
        \sum_{j=1}^{n} A^\prime(i, j) = d_i = \sum_{j=1}^{n} A(i, j)\,.
    \end{equation}
\end{property}
\begin{proof}
    \begin{align}
            \sum_{j} A^\prime(i, j) &= \sum_{l} \sum_{j\in S_l} \frac{d_i}{D_k} A_S(k, l) \frac{d_j}{D_l} \nonumber \\
            &= \sum_{l} \frac{d_i}{D_k} A_S(k, l) \\
            &= d_i \, . \nonumber
    \end{align}
\end{proof}

Figure~\ref{fig:toy_example} demonstrates the difference between these two reconstruction schemes. 
It can be seen that our CR scheme can yield more accurate results and restore the graph topology and adjacency matrix better than uniform reconstruction scheme.

\subsection{Our proposed DPGS}
We use MDL principle to find a good summarization by minimizing the total description length, which assumes that in a communication game, one needs to pass
the shortest encoding bits of the summary graph, and the error
information to the other one for reconstructing the original graph.
Then the objective $L(M, D)$ of our DPGS 
has two parts: description length of summary graph $L(M)$, and error description length $L(D\mid M)$: 
\begin{equation}
    L(M, D) = L(M) + L(D\mid M) \, .\label{equ:mdl}
\end{equation}

To encode the error between original and reconstructed adjacency matrices $A$ and $A^\prime$, 
we use generalized KL-divergence, an instance of Bregman divergence~\cite{dhillon2005generalized} like \cite{henderson2012rolx}:
\begin{equation}\label{equ:error_part}
    \begin{aligned}
    L(D\mid M) &= \KL(A\| A^\prime) \\
    &= \sum_{i, j} A(i, j)\ln \frac{A(i, j)}{A^\prime (i, j)} - A(i, j) + A^\prime (i, j)\, \\
    &= \sum_{i, j} A(i, j)\ln \frac{A(i, j)}{A^\prime (i, j)},
    \end{aligned}
\end{equation}
where the last two terms $-A(i, j) + A^\prime(i, j)$ originate from the first-order term in Bregman divergence.
Due to the degree-preservation property, the last two terms are
cancelled out when summing over $i$ and $j$. 

Moreover, we show that the eigenvalue perturbation are bounded by the reconstruction error $L(D\mid M)$, as stated in the following theorem:
\begin{theorem}[Eigenvalue perturbation]
    Denote the normalized Laplacian matrix of the original graph and the reconstructed graph as $\mathcal{L}$ and $\mathcal{L}^\prime$. Then the total squared error of their eigenvalues (denoted as $\lambda(i)$ and $\lambda^\prime(i)$) are bounded as follows:
    \begin{equation}
        \sum_{i=1}^{n} ( \lambda(i) - \lambda^\prime(i) )^2 \le 2 \cdot L(D\mid M)
    \end{equation}
\end{theorem}
\begin{proof}
    See the Appendix.
\end{proof}

For the encoded length of model, we have:
\begin{equation}\label{equ:model_part}
    L(M) = \LN(n_s) + n \LN(n_s) + \sum_{i=1}^{n} \LN(d_i) + L(A_S)\, ,
\end{equation}
where $\LN$ is the optimal encoding length for positive integers~\cite{rissanen1978modeling}.
The first two term encodes the number of supernodes, and the supernode index to which each node belongs respectively (we use $\LN(n_s)$ for all nodes for simplicity). 
The third term records degrees of nodes in the original graph since the reconstruction needs degree information. 
Since the degree distribution is skewed, the degree encoding length is not large. 
Finally, $L(A_S)$ encodes the adjacency matrix of the summary graph in the following way:
\begin{equation}
    \begin{aligned}
        L(A_S) &= \LN(m_s) + \sum_{i=1}^{m_s} \LN(w_i)  \\
               &+ \BE\left(\frac{n_s(n_s+1)}{2}, m_s\right)\, ,
    \end{aligned}
\end{equation}
where $w_i$ is the weight of superedge $i$, and $\BE(\frac{n_s(n_s+1)}{2}, m_s)$ encodes $m$ superedges' endpoints (there are $\frac{n_s(n_s+1)}{2}$ possible superedges) using \textbf{binomial encoding}, as Equation~\eqref{equ:BE} shows.
In summary, $L(A_S)$ encodes the size, endpoints and weights of superedges.
\begin{equation}\label{equ:BE}
    \BE(a, b) = -b \log_2 \frac{b}{a} - (a-b)\log_2 (1-\frac{b}{a}) \, .
\end{equation}

\subsection{Algorithm}
Our algorithm is based on greedy merging operation. Firstly, each node is initialized as a supernode containing itself alone. Then, the algorithm finds promising supernode pairs $(u, v)$ and merges them together consecutively. Here, the goodness of node pair $gain(u, v)$ is defined as the decrease of the total description length by merging them.

In the ideal case, the algorithm is expected to find the best pairs at each step, which leads to the greatest decrease of the description length. However, this exhaustive search takes $O(|V|^2)$ time for each step, which is time-consuming and not scalable to large graphs. Thus, we must find a way to reduce the search space.

The total description length is comprised of two parts, the model part and the error part. The model part is trivial and hard to analyze quantitatively. Generally, the smaller the $m_s$ and $n_s$ are, the smaller the model length is. 

For the error part, denote the change of error length by merging supernode $S_i$ and $S_j$ as $\Delta L_E(i, j) = L(D\mid M^\prime) - L(D\mid M) $, where $L(D\mid M^\prime)$ and $L(D\mid M)$ refer to the error description length at current step and that after merging them.
\begin{theorem}[Merging Cost]\label{thm:cost}
    $\Delta L_E(i, j) \ge 0$.
\end{theorem}
\begin{proof}
    See the Appendix.
\end{proof}
Theorem \ref{thm:cost} says that merging two nodes never decreases the error part length. This fits intuition since merging two nodes loses information. 
Thus, $\Delta L_E(i, j)$ can be seen as the cost of merging supernode $S_i$ and $S_j$. While the error length increases due to the merging, the model part may decrease, and this leads to the reduction of total description length overall.

By analyzing $\Delta L_E(i, j)$, we have the following observation.
\begin{observation}
    The more common neighbors supernode $S_i$ and $S_j$ has, the more likely merging cost of them is small.
\end{observation}

\begin{algorithm}[t]
    \caption{DPGS}
    \label{alg:DPGS}
    \begin{algorithmic}[1]
        \REQUIRE{$G=(\mathcal{V}, \mathcal{E})$, iteration $T$}
        \ENSURE{$G_S=(\mathcal{V}_S, \mathcal{E}_S, A_S)$}
        \STATE{$G_S \gets G, \mathcal{V}_S \gets \mathcal{V}, \mathcal{E}_S \gets \mathcal{E}$}
        \STATE{$t \gets 0$}
        \WHILE{$t < T$}
            \STATE{$t\gets t+1$}
            \STATE{Update LSH}
            \STATE{Divide supernodes into disjoint groups according to LSH}
            \FOR {each group $g$}
                \STATE{MergeGroup($g$)}
            \ENDFOR
        \ENDWHILE
        \RETURN $G_S$
    \end{algorithmic}
\end{algorithm}

Due to the space limit, the full analysis is placed in the Appendix. This fact provides a guideline for our algorithm that we should merge nodes with many common neighbors. 

The overview of the algorithm is shown in Algorithm~\ref{alg:DPGS}. Basically, we utilized LSH technique~\cite{andoni2006near} to group nodes with similar neighborhoods together, and try merging node pairs separately in each group. This process is repeated for $T$ turns. More specifically, we adjust the parameters of LSH $b$ dynamically. It is known that the bigger parameter $b$ is, the easier two nodes with common neighbors are grouped together. For early iteration, we use small $b$ to guarantee the node pair quality. As the iteration goes on, we increase $b$ gradually to explore more candidate pairs.

\begin{algorithm}[t]
    \caption{MergeGroup}
    \label{alg:group_merge}
    \begin{algorithmic}[1]
        \REQUIRE{$g \subset V_S$} 
        \STATE{$times \gets \log_2 |g|$}
        \STATE{$nskip \gets 0$}
        \WHILE{$nskip < times$ and $|g| \ge 1$}
            \STATE{$pairs \gets$ Sample $\log_2 |g|$ node pairs from $g$}
            \STATE{$u, v \gets \arg\max_{(i, j)\in pairs} gain(i, j)$}
            \IF{$gain(u, v) > 0$}
                \STATE{Merge $u$ and $v$}
                \STATE{$nskip \gets 0$}
            \ELSE
                \STATE{$nskip \gets nskip + 1$}
            \ENDIF
        \ENDWHILE
    \end{algorithmic}
\end{algorithm}

Inspired by SSumM~\cite{lee2020ssumm}, in each group $g$, we sample $\log_2 |g|$ node pairs and merge the one with maximum gain. Here the gain is defined as the reduction of . This process is repeated until the size of group is less than 2, or the algorithm fails to find node pairs decreasing description length for $\log_2 |g|$ times. The full algorithm is described in Algorithm \ref{alg:group_merge}

\subsection{Complexity Analysis}
DPGS scales linearly with the number of edges of input graph, as stated in Theorem~\ref{thm:scalability}.

\begin{theorem}\label{thm:scalability}
    The time complexity of Algorithm~\ref{alg:DPGS} is $O(T\cdot |E|)$.
\end{theorem}
\begin{proof}
    In each iteration, updating LSH costs $O(|E|)$ time. For group merging step, at most $\log^2 |g|$ pairs are sampled, and calculating the merging gain of node pair $(u, v)$ (plus the possible merging step) costs $O(d_u+d_v)$ time. If we limit the size of group not larger than a constant $C$ (for example, 500), the expected running time of merging a group is $O(\sum_{u\in g} d_g)$. Thus, merging all the groups costs $O(\sum_g \sum_{u\in g} d_u) = O(|E|)$. In conclusion, the algorithm runs $T$ iterations, and the total time complexity is $O(T\cdot |E|)$.
\end{proof}

\begin{figure*}[t]
    \centering
    \begin{subfigure}{0.5\textwidth}
        \centering
        {\includegraphics[height=0.02\textheight]{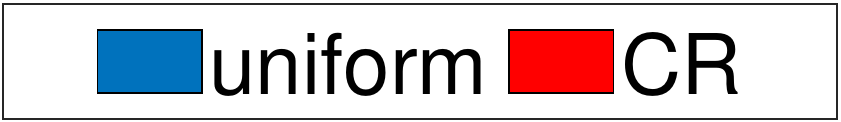}}
    \end{subfigure}
    
    \vspace{1mm}
    \begin{subfigure}{0.22\textwidth}
        \centering
        {\includegraphics[width=\linewidth]{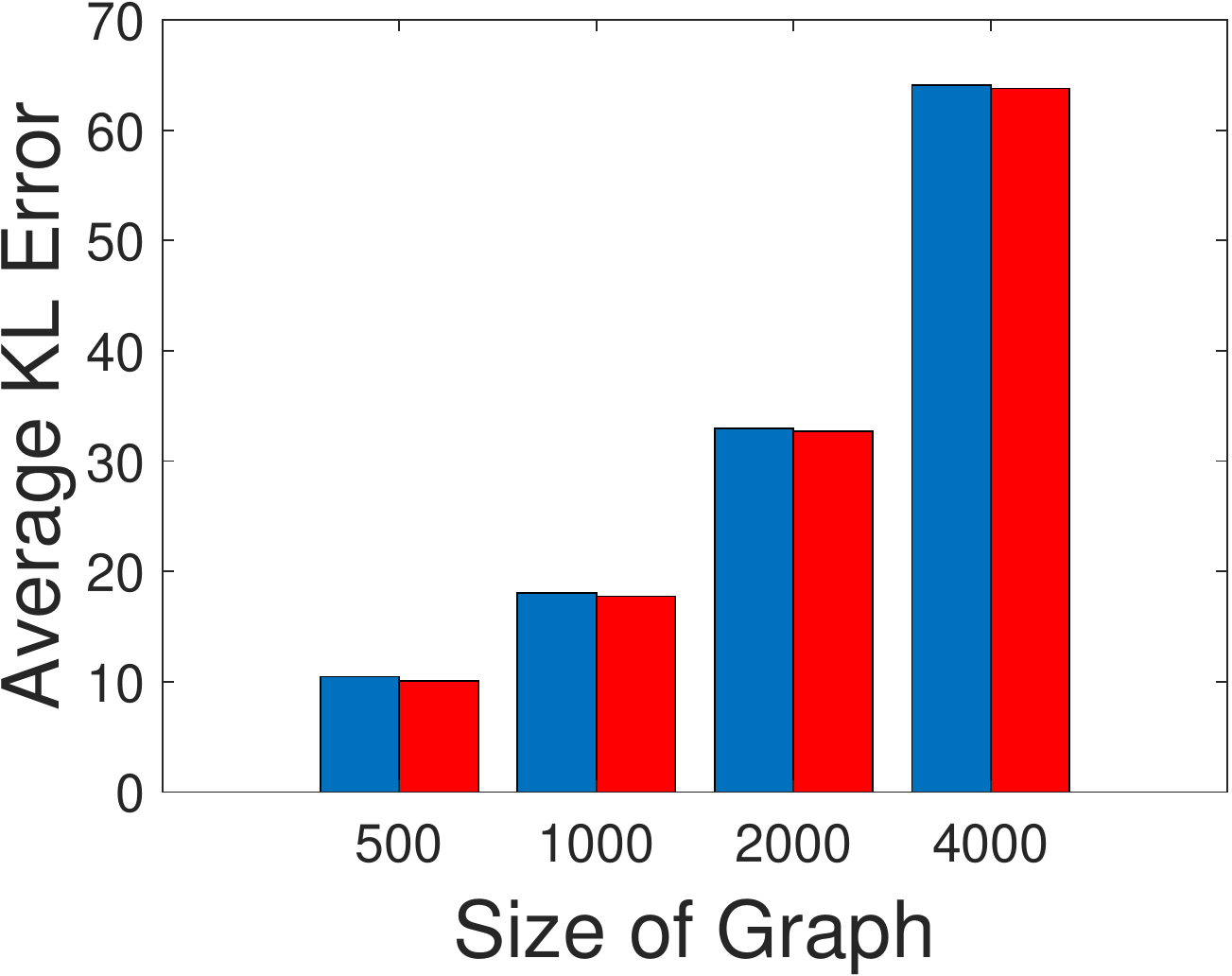}}
        \caption{Erd\H {o}s-R\'enyi ($p=0.02$)}\label{fig:2-5}
    \end{subfigure}
    \quad
    \begin{subfigure}{0.22\textwidth}
        \centering
        {\includegraphics[width=\linewidth]{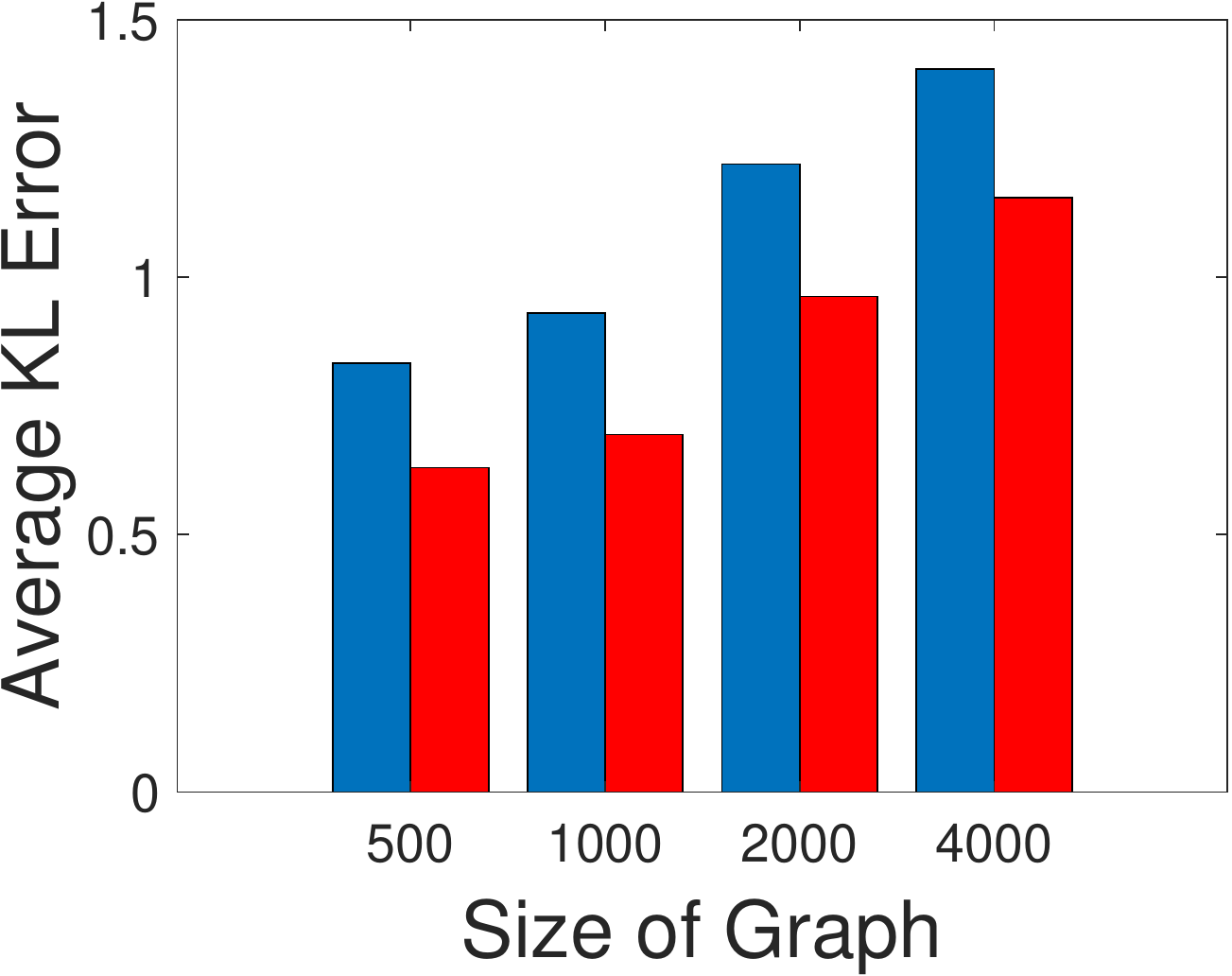}}
        \caption{Power-law ($\alpha=3.0$)}\label{fig:2-6}
    \end{subfigure}
    \quad
    \begin{subfigure}{0.22\textwidth}
        \centering
        {\includegraphics[width=\linewidth]{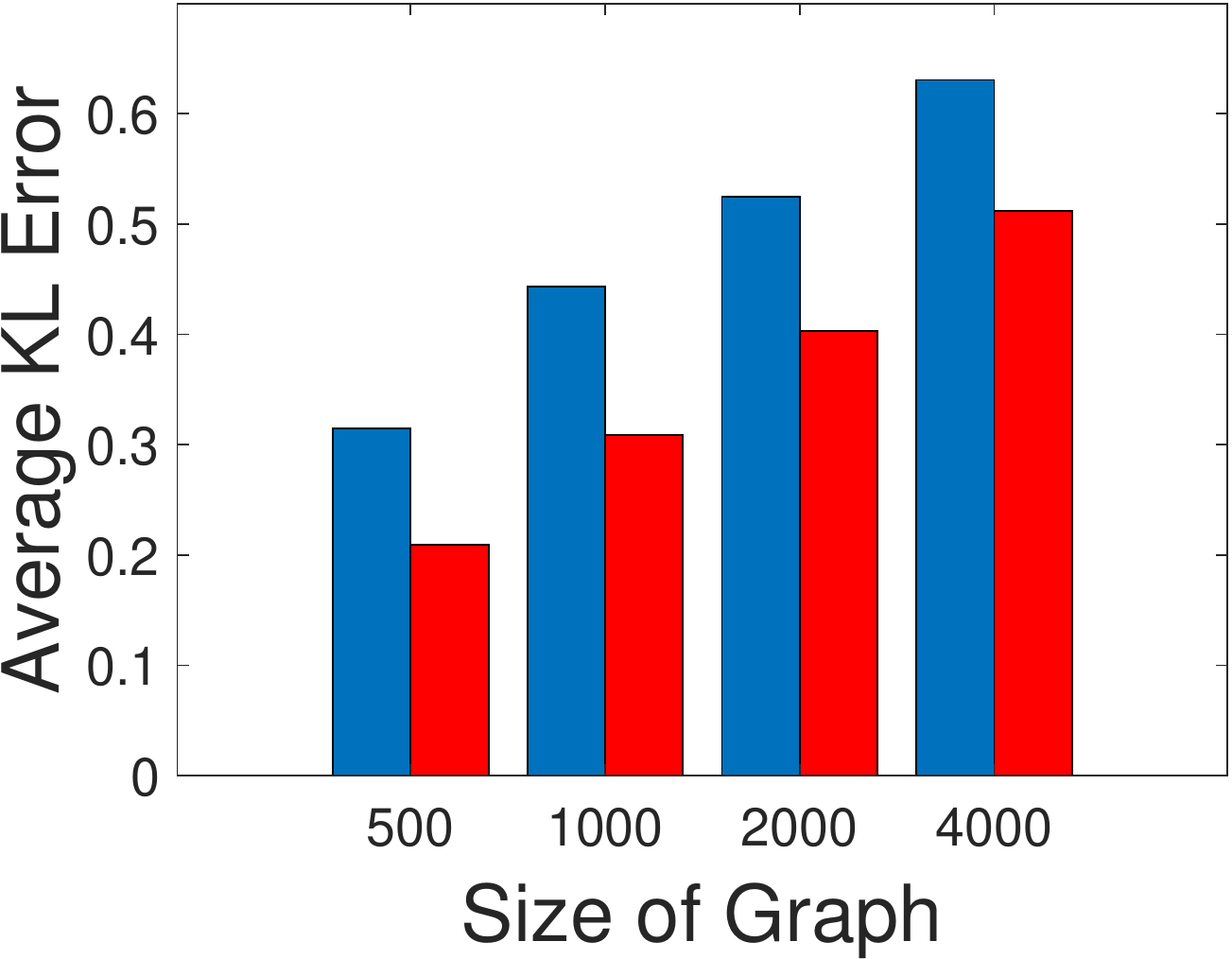}}
        \caption{Power-law ($\alpha=3.5$)}\label{fig:2-7}
    \end{subfigure}
    \quad
    \begin{subfigure}{0.22\textwidth}
        \centering
        {\includegraphics[width=\linewidth]{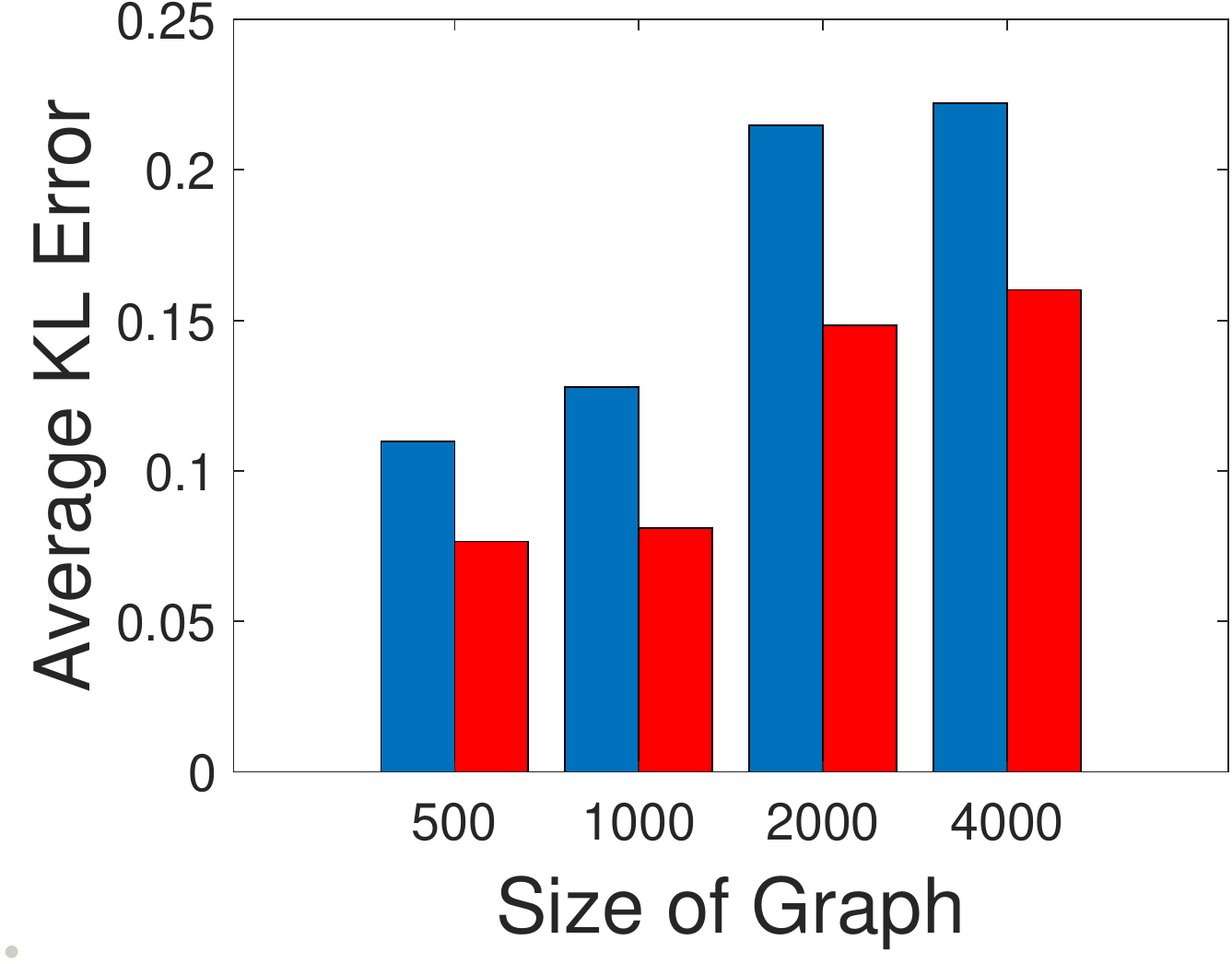}}
        \caption{Power-law ($\alpha=4.0$)}\label{fig:2-8}
    \end{subfigure}
    \caption{Average reconstruction error\protect\footnotemark[1] (KL-divergence) of two reconstruction schemes. Ours achieves lower error when the degrees of nodes are highly skewed (larger $\alpha$ indicates higher skewness), as in many real-world graphs. 
    }
    \label{fig:recon_scheme}
\end{figure*}

\subsection{Compatibility with Existing Methods}
\label{sec:method:comptability}
As a reconstruction scheme is the heart of most summarization models, 
our novel CR reconstruction scheme 
can replace the commonly used uniform reconstruction scheme, and upgrade 
the existing models and then the corresponding algorithms. 
We show in the following two examples: 
\begin{itemize}
    \item \textbf{\kGs(CR):} \kGs\cite{lefevre2010grass} greedily merges (super)node pairs that has the least L1 error increase between the uniformly reconstructed graph and the original one at every step.
    \kGs(CR) upgrade the uniform reconstruction to our CR scheme, i.e.~proportional to node degrees, as well as our KL-divergence encoding
    scheme in formula~\eqref{equ:error_part}).
    \item \textbf{SSumM(CR):} SSumM\cite{lee2020ssumm} greedily merges (super)node pairs and sparsifies (i.e.~drops) (super)edges simultaneously to obtain a sparse summary graph. 
    As SSumM greedily find a summary graph based on MDL, based on uniform
    reconstruction scheme as well. We can smoothly upgrade to our CR scheme
    and error encoding in the optimization objective and algorithm.
\end{itemize}

As such, \kGs(CR) and SSumM(CR) yield degree-preserved summary graphs, 
and inherit the bounding graphs' spectra as our theoretical analysis.



\section{Experiments}
In this section, we design experiments to answer the following questions:
\begin{itemize}
    \item \textbf{Q1. Comparison of reconstruction schemes} Does our CR scheme outperform the uniform reconstruction scheme?
    \item \textbf{Q2. Effectiveness} Does DPGS yield better summarization compared to the baselines on real-world datasets? 
    \item \textbf{Q3. Compatibility} How much can existing graph summarization methods be improved when they are equipped with our reconstruction scheme?
    \item \textbf{Q4. Training GNN on summary graphs} Can we train effective graph neural networks on summary graphs?
    \item \textbf{Q5. Scalability} Does DPGS scale well with the size of the input graph?
\end{itemize}
\begin{table}[ht]
    \caption{Dataset statistics}
    \label{tab:dataset_stats}
    \begin{center}
        \begin{tabular}{cccc}
            \toprule
            Dataset & \#Nodes & \#Edges & Description \\
            \midrule
            ppi & 14,755 & 228,431 & Protein \\
            ppi-large & 56,944 & 818,786 & Protein \\
            soc-Epinions1 & 75,879 & 405,740 & Social \\
            flickr & 89,250 & 449,878 & Social \\
            reddit & 232,965 & 11,606,919 & Social \\
            yelp & 716,847 & 7,335,833 & Social \\
            amazon & 1,569,960 & 132,954,714 & Co-purchase \\
            amazon2m & 2,449,029 & 61,859,140 & Co-purchase \\
            \bottomrule
        \end{tabular}
    \end{center}
\end{table}

\footnotetext[1]{The values are normalized by the size of graph, i.e.~$|V|$.}
\subsection{Q1. Comparison of reconstruction schemes}
We generate synthetic data using the following random graph models:
\begin{itemize}
    \item $G(n, p)$ model (also known as Erd\H {o}s-R\'enyi model)~\cite{erdos59a}, with connection probability $p=0.02$.
    \item Random graph model with power-law degree distribution, with parameter $\alpha=3.0$, $3.5$, and $4.0$ 
\end{itemize}
The main difference between these two models is that the degrees of nodes in Erd\H {o}s-R\'enyi graphs are nearly uniform, while those in power-law graphs are highly skewed (larger $\alpha$ implies higher skewness).

We compare our CR scheme with the uniform reconstruction scheme using the summary 
graphs obtained by SSumM \cite{lee2020ssumm}. 
Note that we fix the optimization method to fairly compare the two reconstruction schemes.
As shown in Figure~\ref{fig:recon_scheme}, our CR scheme achieves lower reconstruction error (i.e. KL-divergence error) than the uniform reconstruction scheme.
The margin of improvement is significantly large when the degree distribution is highly skewed, as in many real-world graphs.
Simply put, our CR scheme has the superiority 
to give better summarization of real-world graphs.

%

\begin{figure*}[t]
    \centering
    \begin{subfigure}{0.5\textwidth}
        \centering
        {\includegraphics[height=0.02\textheight]{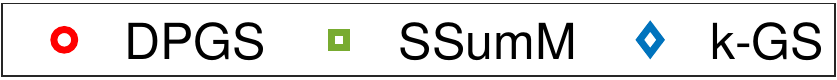}}
    \end{subfigure}
    
    \vspace{1mm}
    \begin{subfigure}{0.21\textwidth}
        \centering
        {\includegraphics[width=\linewidth]{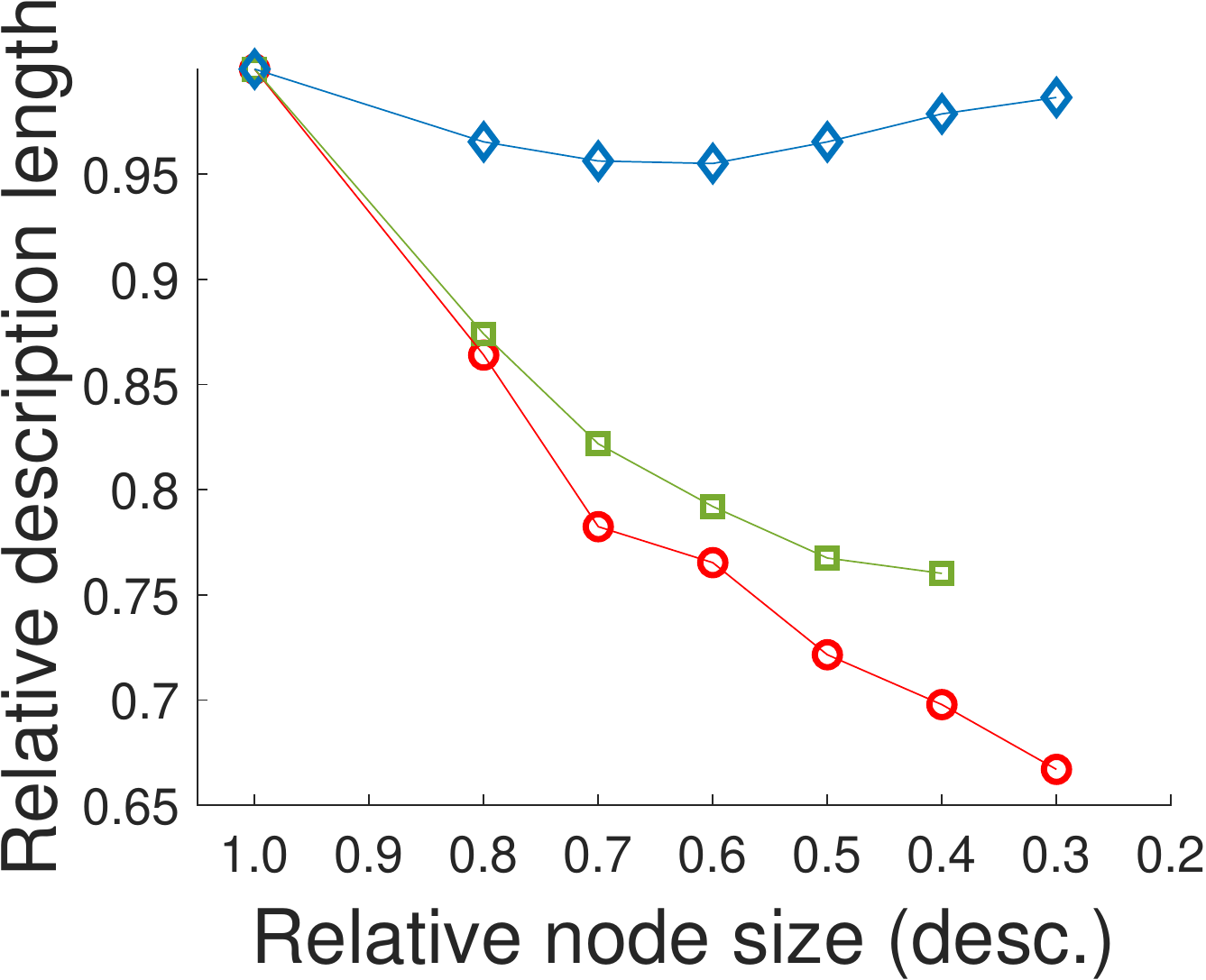}}
        \caption{ppi}\label{fig:3-1}
    \end{subfigure}
    \quad
    \begin{subfigure}{0.21\textwidth}
        \centering
        {\includegraphics[width=\linewidth]{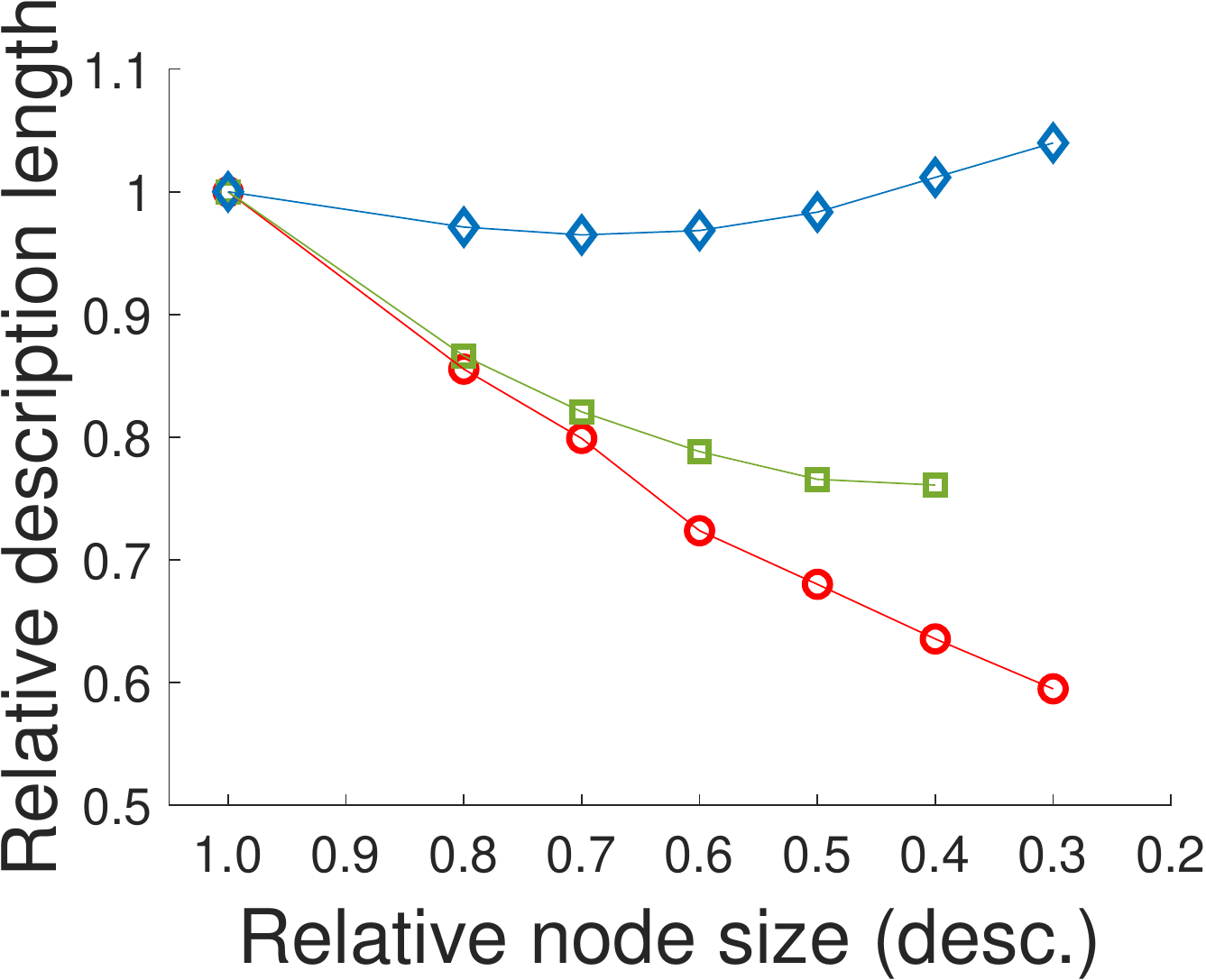}}
        \caption{ppi-large}\label{fig:3-2}
    \end{subfigure}
    \quad
    \begin{subfigure}{0.21\textwidth}
        \centering
        {\includegraphics[width=\linewidth]{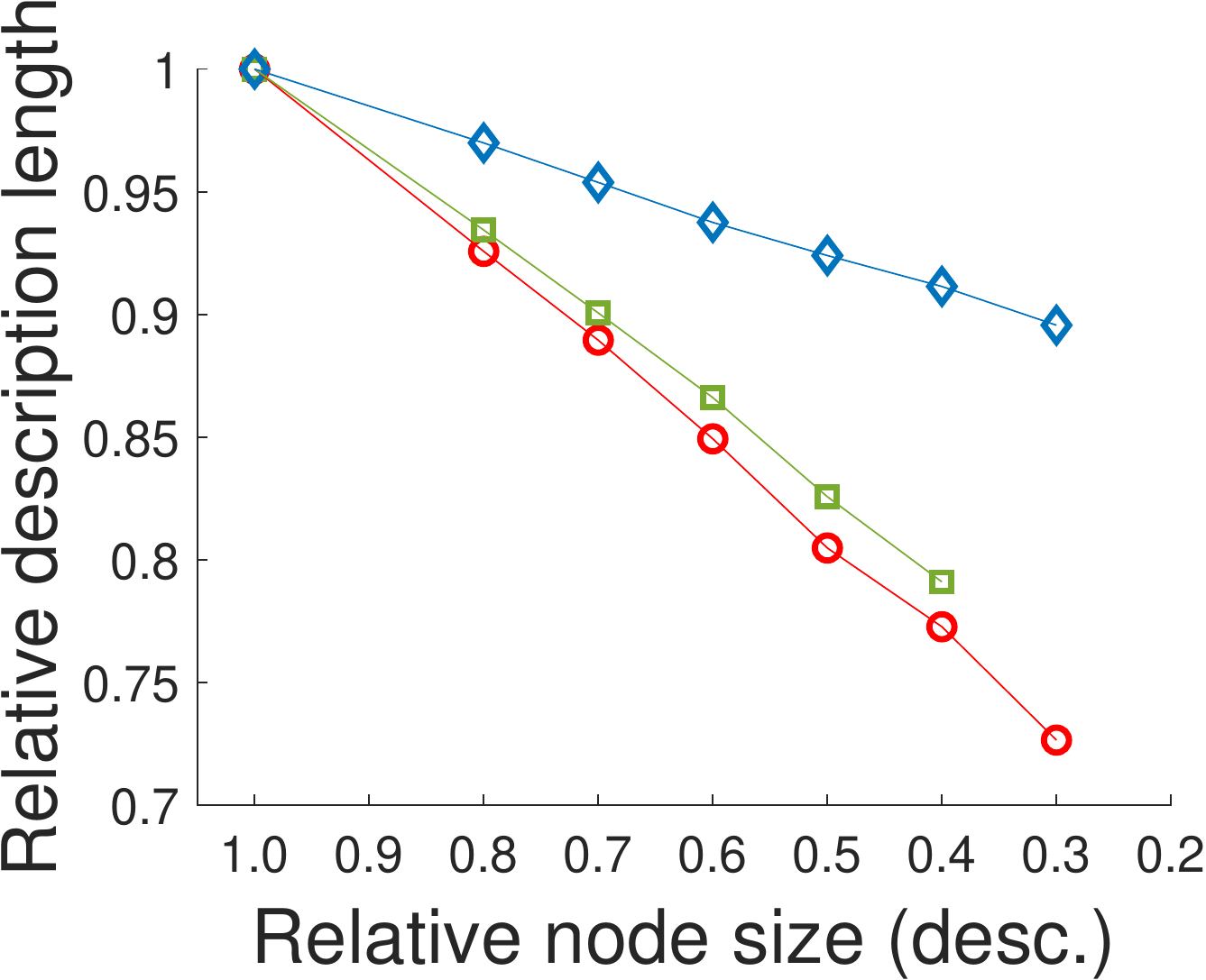}}
        \caption{soc-Epinions1}\label{fig:3-3}
    \end{subfigure}
    \quad
    \begin{subfigure}{0.21\textwidth}
        \centering
        {\includegraphics[width=\linewidth]{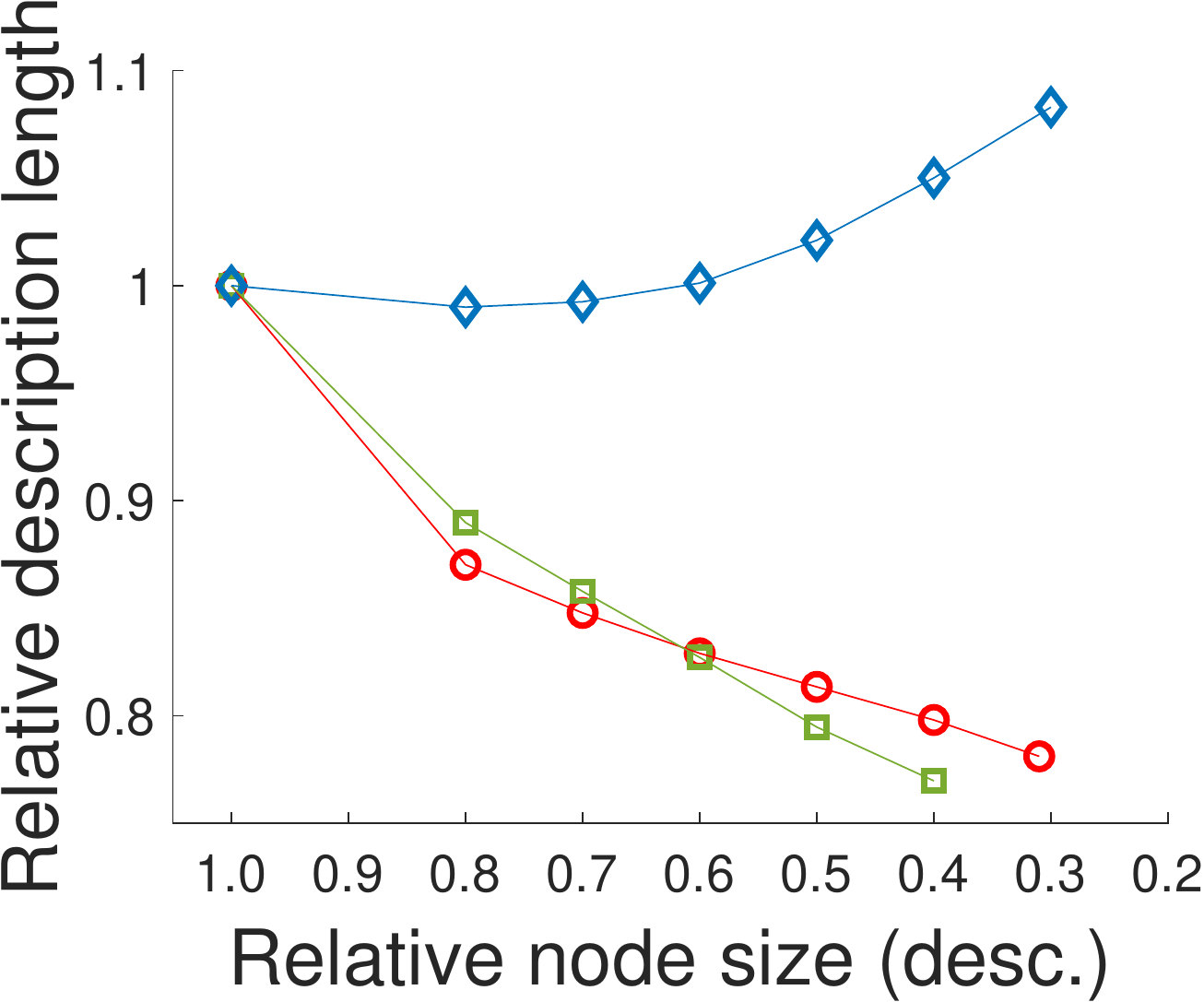}}
        \caption{flickr}\label{fig:3-4}
    \end{subfigure}
    
    \begin{subfigure}{0.21\textwidth}
        \centering
        {\includegraphics[width=\linewidth]{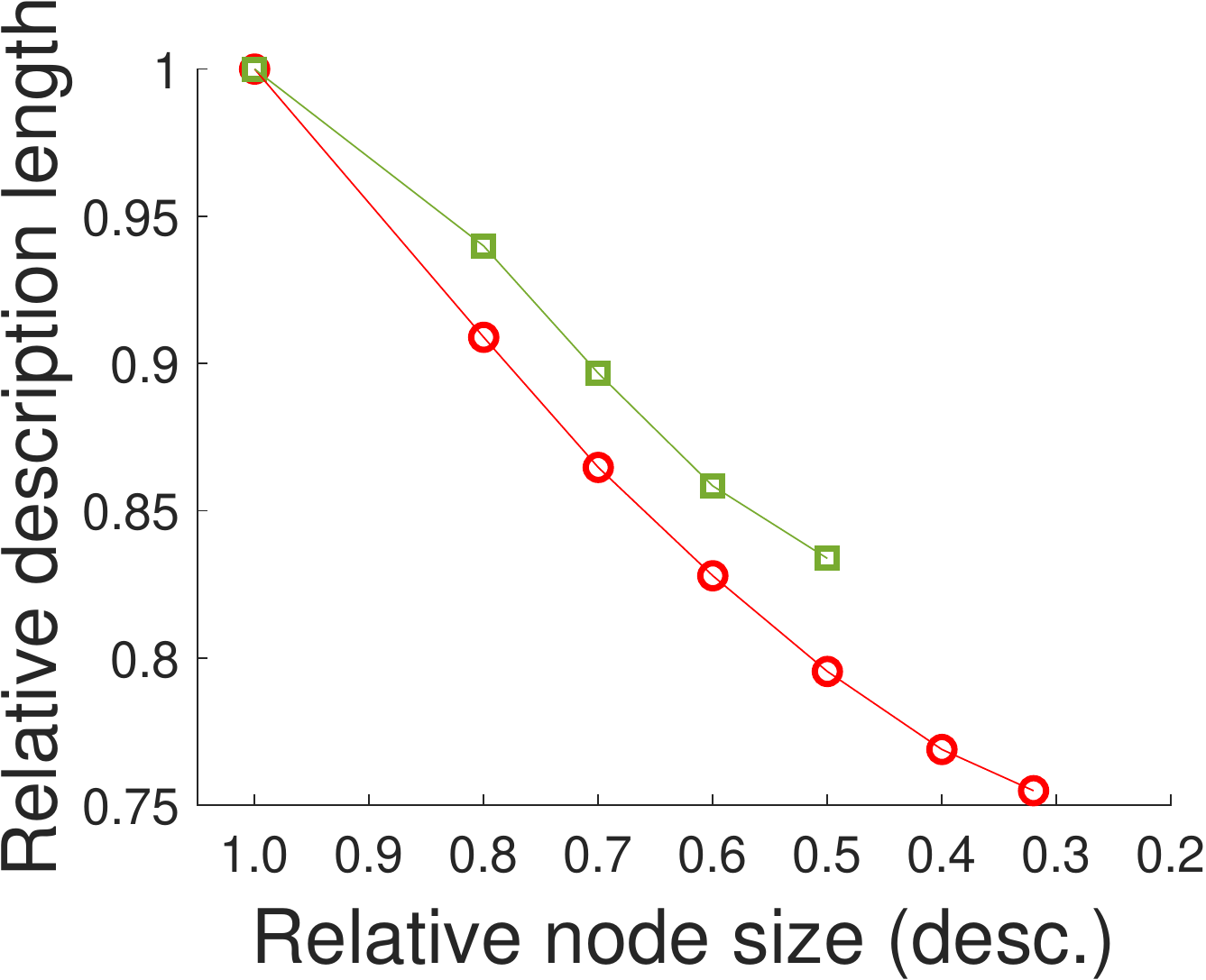}}
        \caption{reddit}\label{fig:3-5}
    \end{subfigure}
    \quad
    \begin{subfigure}{0.21\textwidth}
        \centering
        {\includegraphics[width=\linewidth]{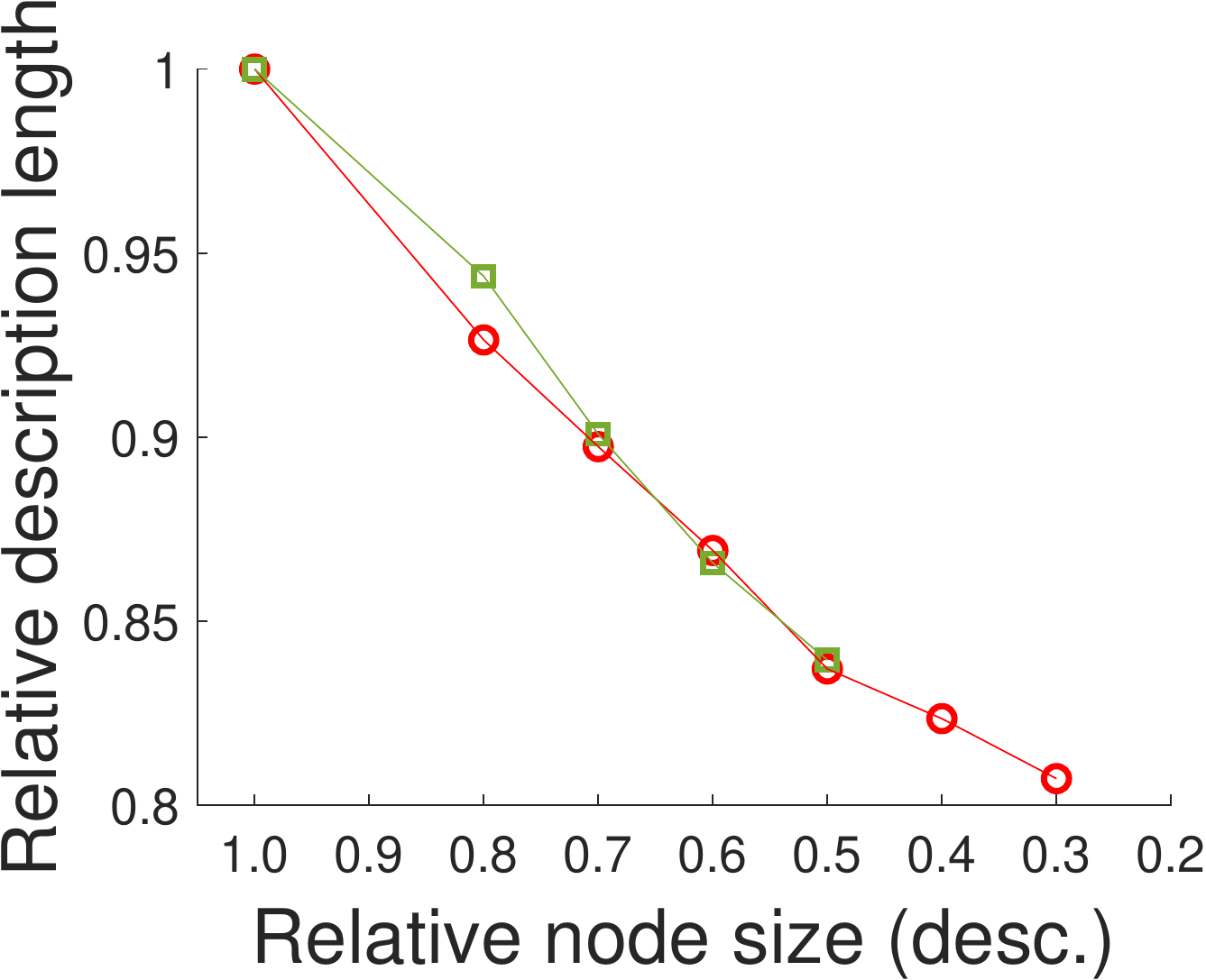}}
        \caption{yelp}\label{fig:3-6}
    \end{subfigure}
    \quad
    \begin{subfigure}{0.21\textwidth}
        \centering
        {\includegraphics[width=\linewidth]{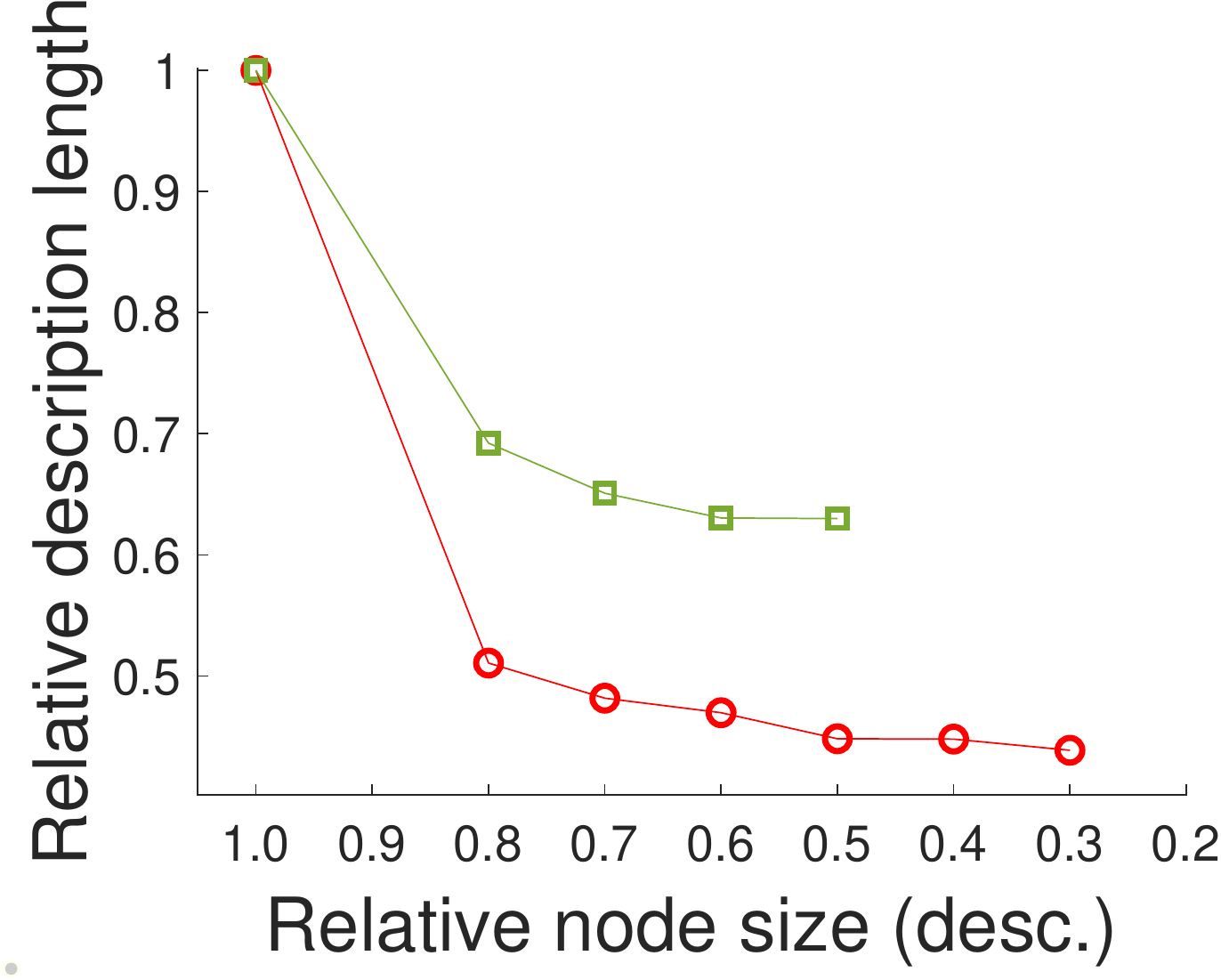}}
        \caption{amazon}\label{fig:3-7}
    \end{subfigure}
    \quad
    \begin{subfigure}{0.21\textwidth}
        \centering
        {\includegraphics[width=\linewidth]{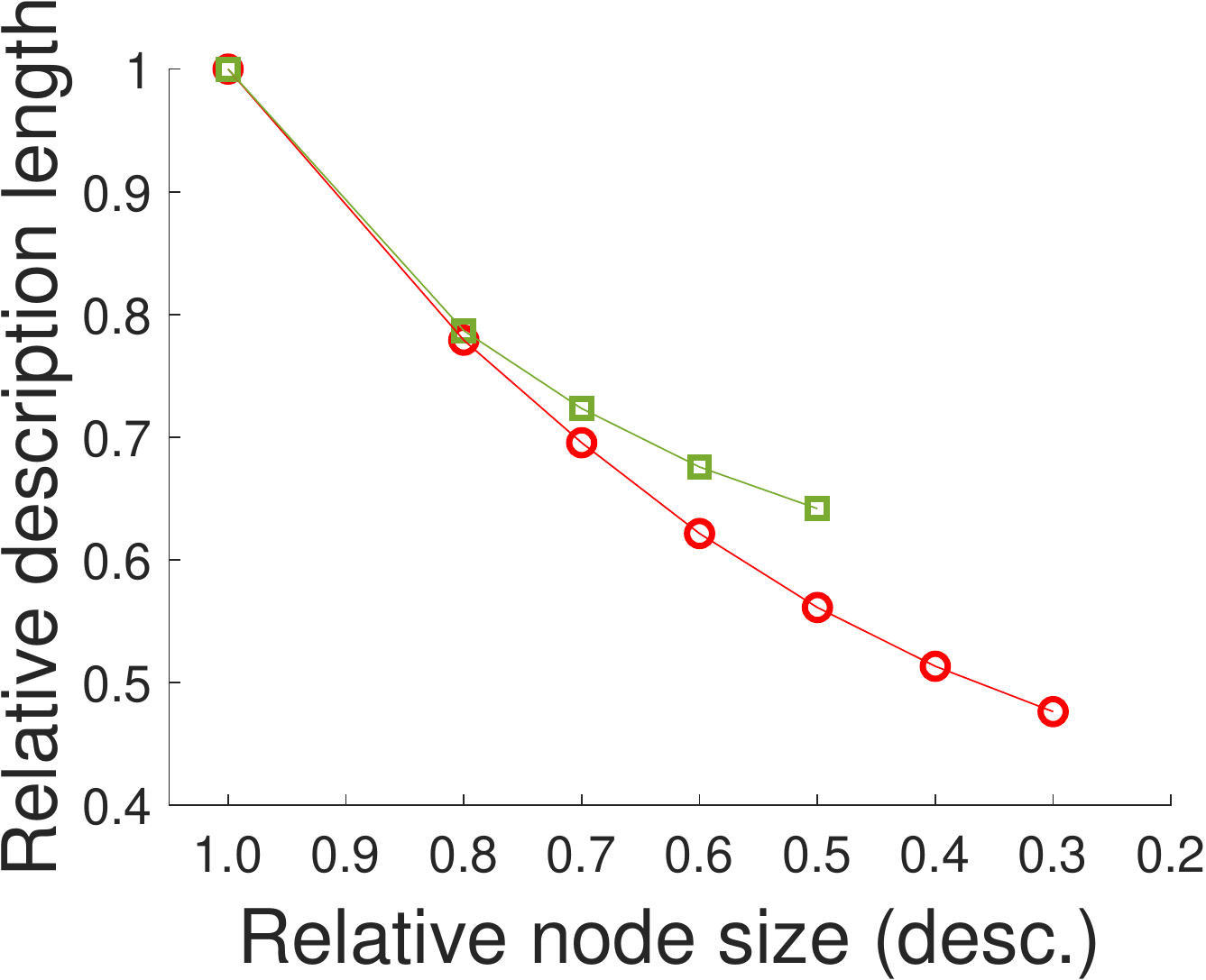}}
        \caption{amazon2M}\label{fig:3-8}
    \end{subfigure}
    \caption{Relative description length\protect\footnotemark[3]  (the smaller, the better) of different methods on 8 datasets. DPGS yields the best summarization result on most datasets. Due to the high complexity of \kGs, it fails on the large datasets in the second row.
    }
    \label{fig:compression_ratio}
\end{figure*}

\subsection{Q2. Effectiveness}
We compare our DPGS approach with \kGs \cite{lefevre2010grass} and SSumM \cite{lee2020ssumm} on eight real-world datasets in Table \ref{tab:dataset_stats}.

We implement \kGs in C++ and adopt the \textsc{SamplePairs} strategy for summarizing large graphs. For SSumM, we use the open-sourced code\footnote[2]{\url{https://github.com/KyuhanLee/SSumM}}. The iteration turn is set to 30. In our DPGS, the iteration turn has the same setting.  
The number of LSH's bands grows gradually with the iteration turn, and the minimum and maximum number is set to 3 and 8 respectively.

We summarize the graphs to different fraction of node size (from 80\% to 20\%),
i.e.~reducing the graph size, and compare the relative description lengths\protect\footnotemark[3] of different methods. 
The results are shown in Figure \ref{fig:compression_ratio}.
It can be seen that the proposed DPGS algorithm can achieve better summarization in most of the datasets, except a small one, flickr, in which we still have very close performance. However, in the other datasets, our DPGS algorithm can achieve up to 28.85\% improvement on relative description length.

\footnotetext[3]{defined as $\sfrac{(L(M) + L(D\mid M))}{ L(D)}$, where $L(D)$ 
is encoding length of the original graph using the same way as encoding the summary graph.}

\subsection{Q3. Compatibility}

In this experiment, we show how our CR scheme 
can improve the existing graph summarization methods. 
We compare \kGs and SSumM, which are based on the uniform reconstruction scheme, with the upgraded \kGs(CR) and SSumM(CR) as described in subsection~\ref{sec:method:comptability}.
Note that the algorithms for the upgraded models search summary graphs by
optimizing the new objective functions, which is different from just
reconstructing a given summary graph with configuration-based scheme in experiments Q1.
Since \kGs searches for the best summary graph subject to a given target number of supernodes,
we measure the KL-divergence error on average while changing the target number
of supernodes (from 10\% to 90\% of the number of nodes in the original input graph).
On the other hand, SSumM searches for the best summary graph subject to a given target size in-bits. Thus, we measure the average KL-divergence error while changing the target size in bits (from 10\% to 80\% of the original input graph size).\footnote[4]{The KL divergence error (i.e., Equation~\eqref{equ:error_part}) is not defined for dropped superedges, 
which may contain $(i,j)\in \mathcal{E}$ such that $A^\prime (i, j)=0$ and $A(i,j)\neq 0$. Thus, we add the correction of each edge belonging to each dropped superedge to the model cost, and we increase the description length (and thus relative size) accordingly. 
} 

\begin{figure*}[t]
    \centering
    \begin{subfigure}{0.5\textwidth}
        \centering
        {\includegraphics[height=0.02\textheight]{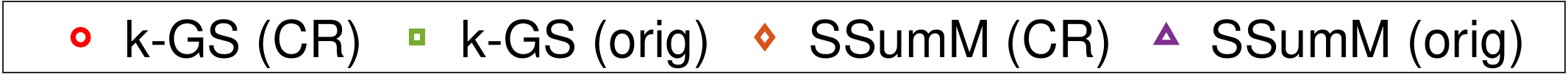}}
    \end{subfigure}
    
    \vspace{1mm}
    \begin{subfigure}{0.25\textwidth}
        \centering
        {\includegraphics[width=\linewidth]{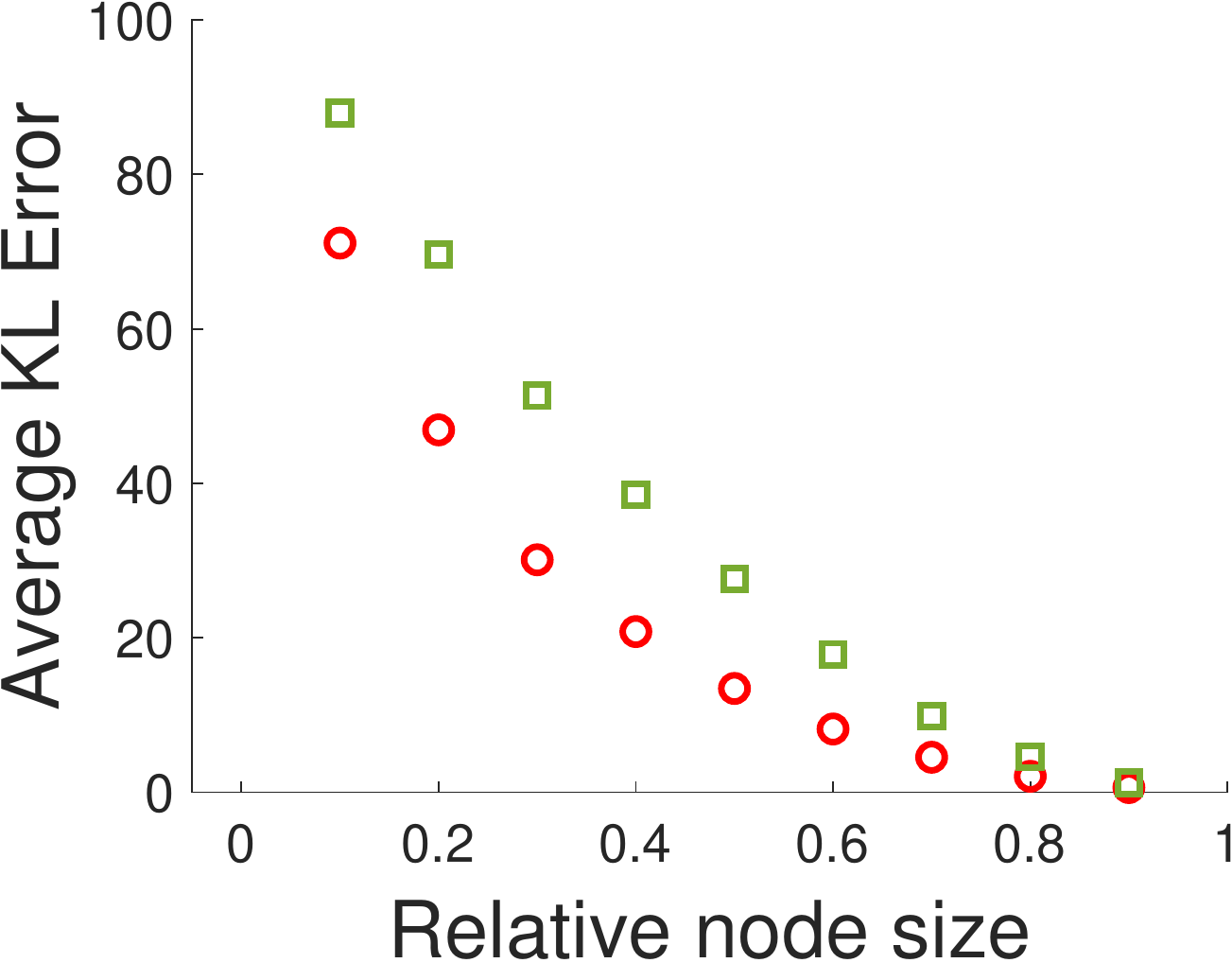}}
        \caption{ppi}\label{fig:4-1}
    \end{subfigure}
    \quad
    \begin{subfigure}{0.25\textwidth}
        \centering
        {\includegraphics[width=\linewidth]{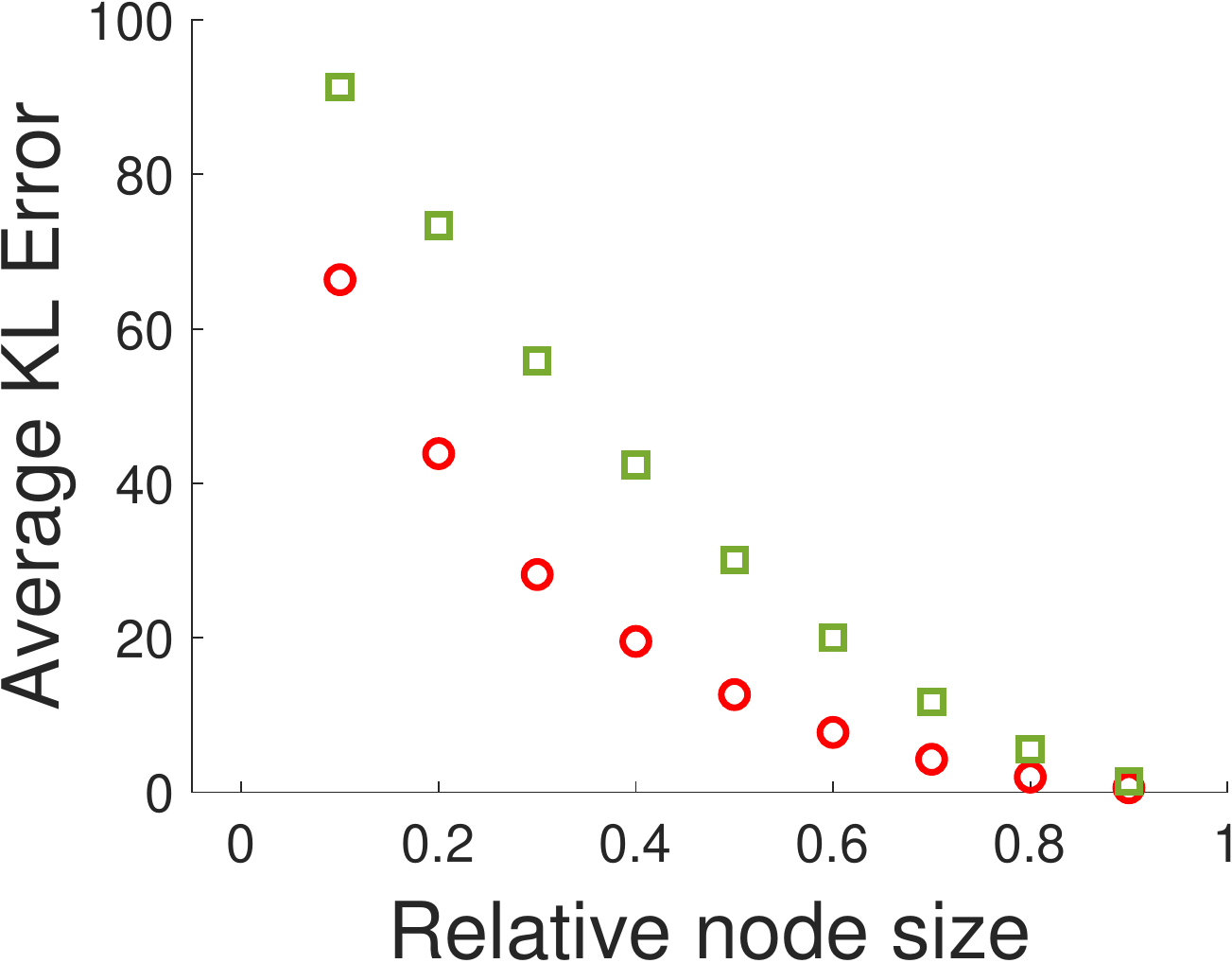}}
        \caption{ppi-large}\label{fig:4-2}
    \end{subfigure}
    \quad
    \begin{subfigure}{0.25\textwidth}
        \centering
        {\includegraphics[width=\linewidth]{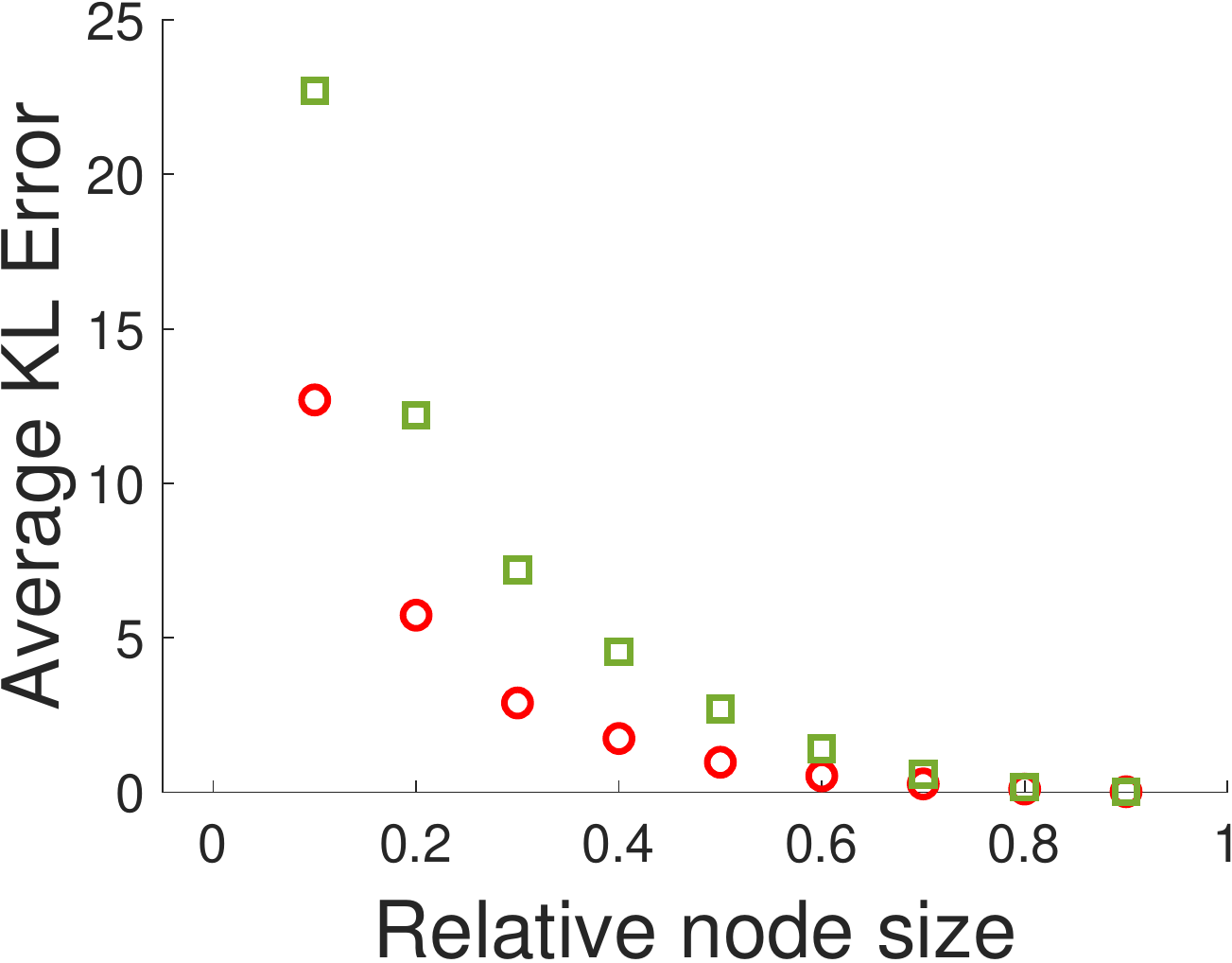}}
        \caption{soc-Epinions1}\label{fig:4-3}
    \end{subfigure}
    
    \begin{subfigure}{0.25\textwidth}
        \centering
        {\includegraphics[width=\linewidth]{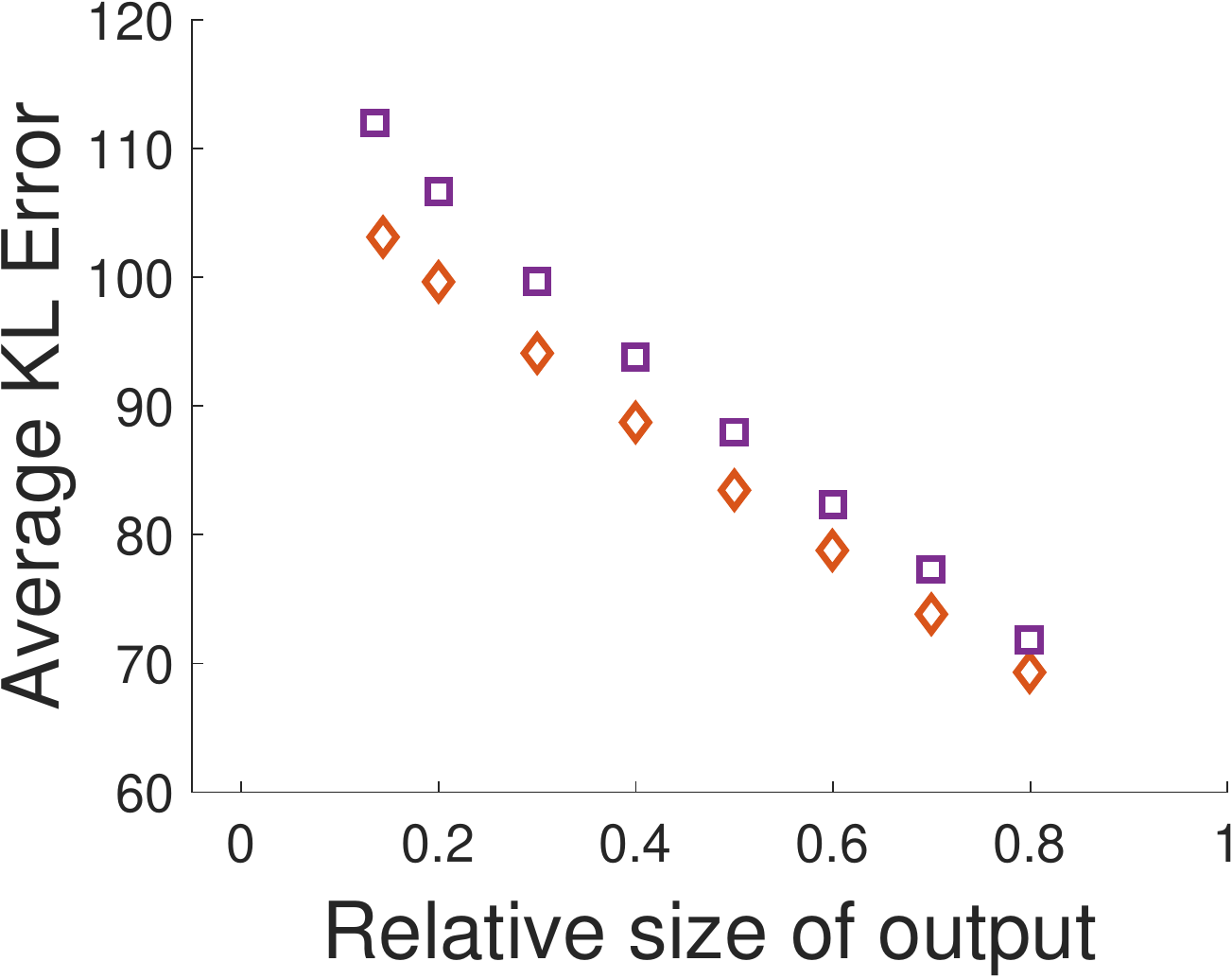}}
        \caption{ppi}\label{fig:4-5}
    \end{subfigure}
    \quad
    \begin{subfigure}{0.25\textwidth}
        \centering
        {\includegraphics[width=\linewidth]{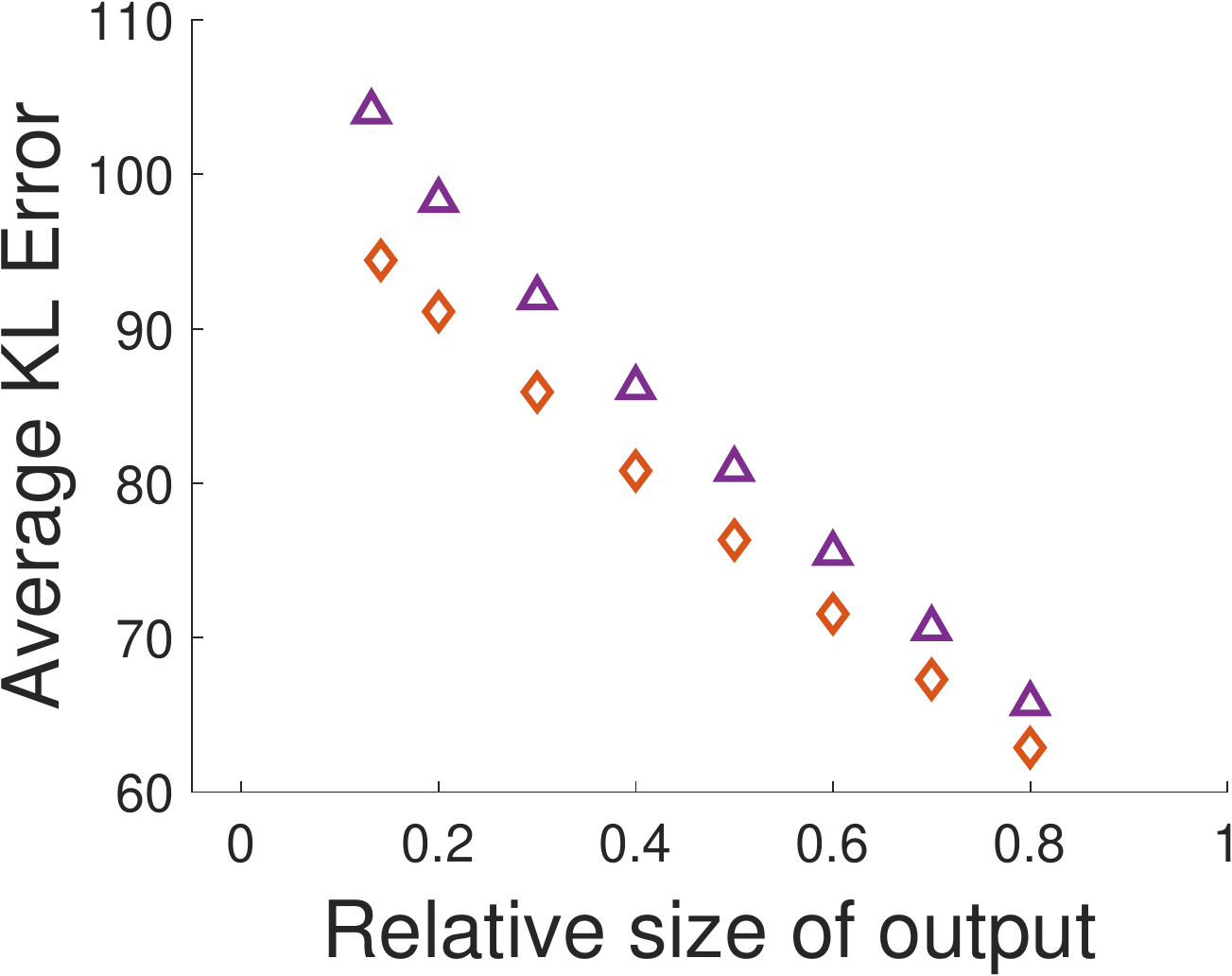}}
        \caption{ppi-large}\label{fig:4-6}
    \end{subfigure}
    \quad
    \begin{subfigure}{0.25\textwidth}
        \centering
        {\includegraphics[width=\linewidth]{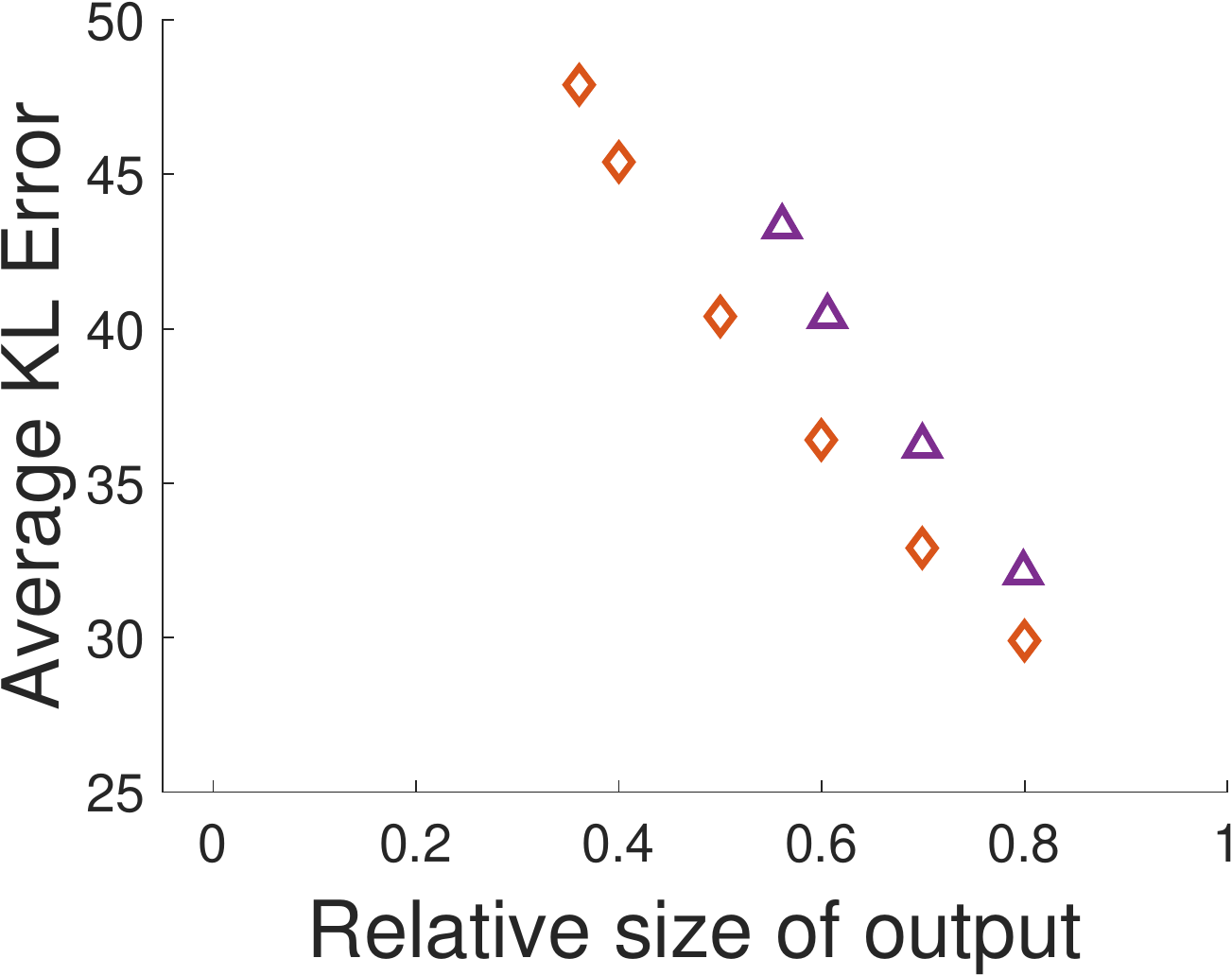}}
        \caption{soc-Epinions1}\label{fig:4-7}
    \end{subfigure}
    \caption{
    Our CR reconstruction scheme improves \kGs and SSumM, which consistently yield better summary graphs, i.e.~smaller reconstruction errors\protect\footnotemark[5], than the original methods.
    Larger improving margins can be seen when reducing to a smaller summary graphs.
    }
    \label{fig:compatibility}
\end{figure*}

As seen in Figure~\ref{fig:compatibility}, the variants with our CR scheme consistently yield more accurate summaries than the original methods based on the uniform reconstruction scheme in all three datasets. For example, \kGs(CR) gives a 2.8$\times$ more accurate summary graph with the same number of supernodes than \kGs(orig) in the ppi-large dataset. Moreover, SSumM(CR) gives a summary graph with 1.2$\times$ smaller KL-divergence error but smaller sizes than SSumM(orig) in the soc-Epinions1 dataset.
Simply put, our CR scheme as the heart of the summarization models 
can help the existing methods to find a better solution.

\subsection{Q4. Training GNNs on summary graphs}
One important application of graph summarization is to accelerate graph mining algorithms, and graph neural networks (GNNs) are one of the most memory-consuming and time-consuming methods currently. 
In the ideal case, we can reduce the running time and required memory via graph summarization, without sacrificing performance much. Thus, we design experiments to check how our summarization method affects the performance of GNN models.
\begin{figure}[htbp]
    \centering
    \includegraphics[width=0.7\columnwidth]{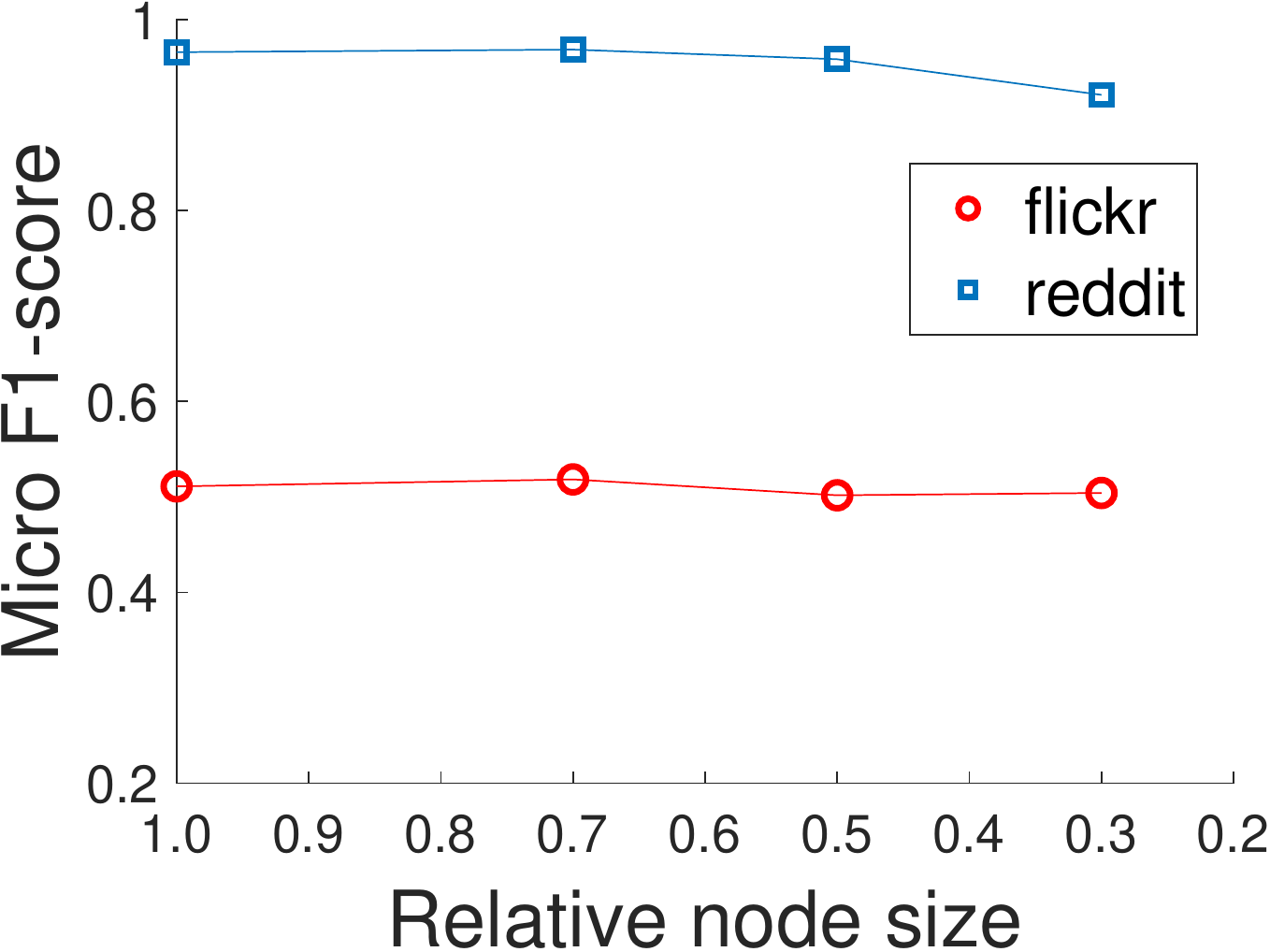}
    \caption{Micro F1-scores of node classification task using models training on summary graphs with different fraction of sizes. 
    }
    \label{fig:GNN}
\end{figure}

Specifically, we choose the task of node classification to test the performance.
Since the graphs we are dealing with are large, we use GraphSAINT~\cite{zeng2020graphsaint}, a scalable GNN model based on subgraph sampling. 

The procedure is as follows: We train GraphSAINT models on summary graphs, keep models' parameters, and test the node classification performance on original graphs. 
\footnotetext[5]{The values are normalized by the size of graph, i.e. $|V|$.}
Features of supernodes are obtained by simple aggregation (i.e., sum) of features of inside nodes, and class labels of supernodes are determined by a majority vote of inside nodes. 
We follow the original GraphSAINT paper's configurations and use the random walk sampler, which achieves the best performance in most datasets. 

We perform experiments on two datasets: flickr and reddit. The micro F1-scores are shown in Figure~\ref{fig:GNN}. Surprisingly, summarization does not harm the performance, even when the size of the summary graphs is 30\% of the original ones.

\begin{figure}[thbp]
    \centering
    \includegraphics[width=0.7\columnwidth]{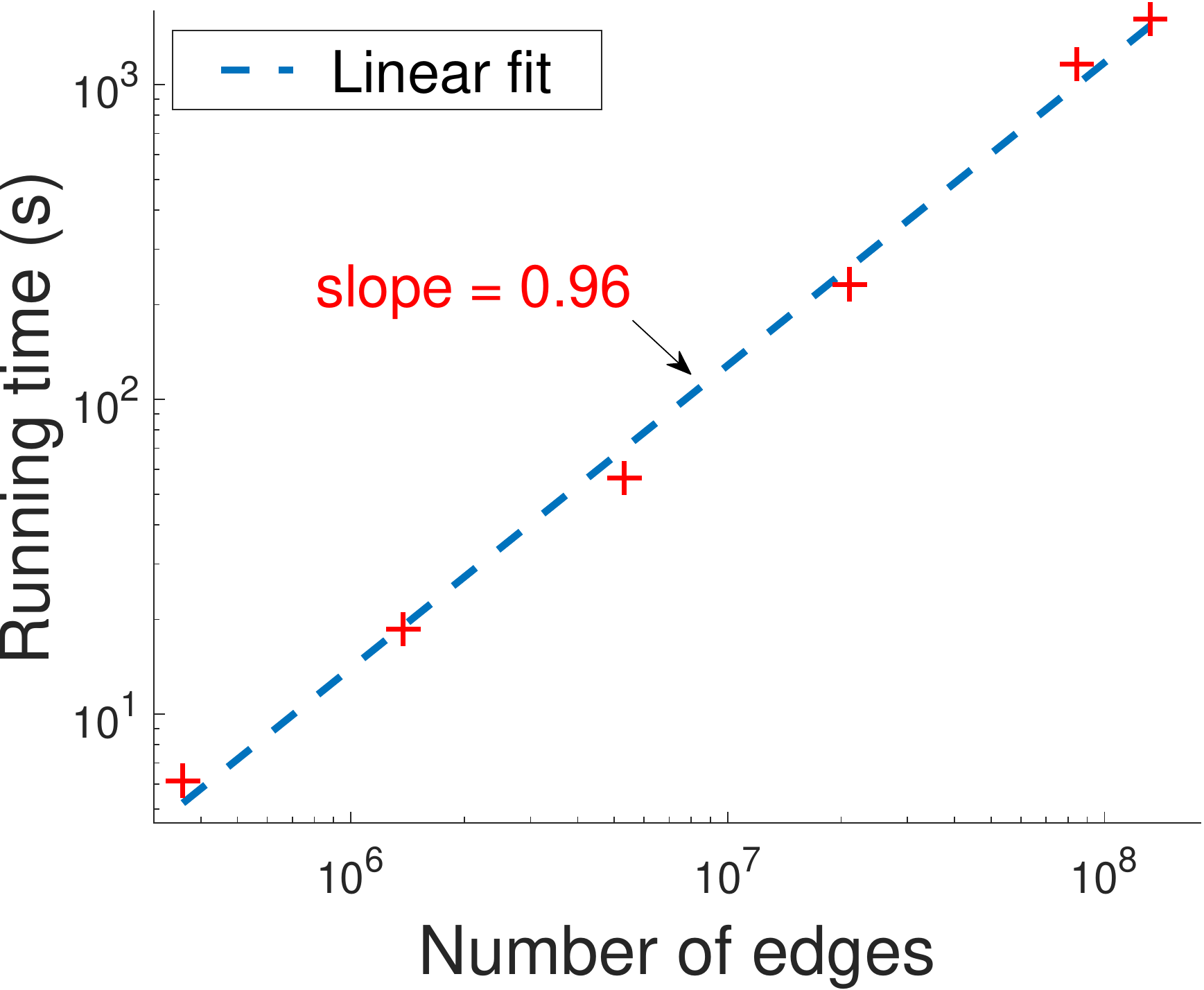}
    \caption{DPGS is scalable. The running time scales linearly with the number of edges.}
    \label{fig:scalability}
\end{figure}

\subsection{Q5. Scalability}
We evaluate the scalability of our method on the largest amazon dataset, which contains 1,569,960 nodes and 132,169,374 edges. We run DPGS algorithm on a number of graphs that are obtained from the original dataset by randomly sampling different numbers of nodes. As seen in Figure~\ref{fig:scalability}, our proposed DPGS algorithm \textbf{scales linearly with the number of edges}.

\section{Conclusion}
In this work, we present a graph summarization algorithm \method using a novel configuration-based reconstruction scheme, and we theoretically show that it can bound the perturbation of graph spectrum.
The proposed reconstruction scheme can be easily applied to existing graph summarization methods to improve their summary result.
Based on the MDL principle, we design an efficient and effective graph summarization algorithm.
Extensive experiments on both synthetic and real-world datasets show that our \method yields more compact results compared to state-of-the-art methods.
Moreover, we show that our summary graph can be effectively used to train GNN models with comparable performance using much less computational resources.


\bibliographystyle{siamplain}
\bibliography{ref_short}

\end{document}


\maketitle
\appendix\section{Analysis of $\Delta L_E (i, j)$}
Assumed that the input graphs are simple and undirected, the error description length $L(D\mid M)$ is:
\begin{equation}\label{equ:error_part2}
    \begin{aligned}
        L(D\mid M) & = \sum_{(i, j)\in E} \ln \frac{1}{\frac{d_i}{D_k} A_S(k, l) \frac{d_j}{D_l}} \\
        & = \sum_{(i, j)\in E} \ln \frac{D_k D_l}{A_S(k, l)} - \sum_{(i, j)\in E} \ln(d_i d_j)\, .
    \end{aligned}
\end{equation}
Since the second term only depends on the input graph, we only consider the first term in the following.
\begin{equation}\label{equ:error_part3}
    \begin{aligned}
        L(D\mid M) &=  \sum_{(i, j)\in E} \ln \frac{D_k D_l}{A_S(k, l)} \\
        &= 2 \sum_{k=1}^{n_s} D_k \ln D_k - \sum_{k=1}^{n_s}\sum_{l=1}^{n_s} A_S(k, l) \ln A_S(k, l) \,.
    \end{aligned}
\end{equation}

Suppose the current supernode set is $\mathcal{S}$. After merging supernodes $S_i$ and $S_j$ into a new supernode $S_k$, we get a new supernode set $\mathcal{S}^\prime = \mathcal{S} \setminus \{S_i, S_j\} \cup \{S_k\}$. The connectivity of $S_k$ is the aggregation of $S_i$ and $S_j$, that is,
\begin{equation}
    A_S(k, l) = \begin{cases}
        A_S(i, l) + A_S(j, l)\, & l \ne k \\
        A_S(i, i) + A_S(j, j) + 2A_S(i, j)\, & l = k \, .
    \end{cases}
\end{equation}

Now we expand the $\Delta L_E(i, j)$ as follow (we define $f(x) = x\ln x (x > 0)$ and $f(0) = 0$ for simplicity):
\begin{equation}\label{equ:merge_cost}
    \begin{aligned}
    \Delta L_E(i, j) &= L(D\mid M^\prime) - L(D\mid M) \\
                &= 2 (f(D_i+D_j) - f(D_i) - f(D_j)) \\
                &+ 2 \sum_{l\in \mathcal{S}, l\neq i, j} [f(A_S(i, l)) + f(A_S(j, l))] \\
                &- 2 \sum_{l\in \mathcal{S}^\prime, l\neq k} f(A_S(k, l)) \\
                &+ f(A_S(i, i)) + f(A_S(i, j)) + 2 f(A_S(i, j)) \\
                &- f(A_S(k, k)) \, .
    \end{aligned}
\end{equation}
Note that if $S_l$ is not the common neighbor of supernode $S_i$ and $S_j$, it makes no contribution to $\Delta L_E(i, j)$. Each common neighbors makes positive contribution to $\Delta L_E(i, j)$ (since $f(x+y) > f(x)+f(y)$ for $x, y \ge 1$). Thus, the more common neighbors supernodes $S_i$ and $S_j$ have, the more likely the merging cost $\Delta L_E(i, j)$ is small.

\section{PROOFS}
\subsection{Proof of Theorem 3.1}
\begin{proof}
    Denote the normalized Laplacian matrix of the original graph and the reconstructed graph as $\mathcal{L}$ and $\mathcal{L}^\prime$. 
    By~\cite{weyl_asymptotische_1912}, the squared error between eigenvalues of $\mathcal{L}$ and $\mathcal{L}^\prime$ are bounded by:
    \begin{equation}
        \sum_{i=1}^{n} ( \lambda(i) - \lambda^\prime(i) )^2 \le \| \mathcal{L} - \mathcal{L}^\prime \|_F^2
    \end{equation}
    
    \begin{align*}
        \| \mathcal{L} - \mathcal{L}^\prime \|_F^2 &= \sum_{i=1}^{n} \sum_{j=1}^{n} | \mathcal{L}(i, j) - \mathcal{L}^{\prime}(i, j) |^2 \\
        &= \sum_{i=1}^{n} \sum_{j=1}^{n} \left( \frac{A(i, j)}{\sqrt{d_i d_j}} - \frac{A^\prime(i, j)}{\sqrt{d_i d_j}} \right)^2 \\
        &\le \sum_{i=1}^{n} \sum_{j=1}^{n} \left( \frac{A(i, j)}{\sqrt{d_i}} - \frac{A^\prime(i, j)}{\sqrt{d_i}} \right)^2 \\
        &= \sum_{i=1}^{n} \sum_{j=1}^{n} d_i \left( \frac{A(i, j)}{d_i} - \frac{A^\prime(i, j)}{d_i} \right)^2 \\
    \end{align*}
    
    Denote the normalized $i$-th row of $A$ and $A^\prime$ as $\tilde{A}(i)$ and $\tilde{A}^\prime(i)$, then $\tilde{A}(i)$ and $\tilde{A}^\prime(i)$ can be seen as two distributions. Moreover, with the following inequality~\cite{thomas2006elements}:
    \begin{equation}\label{equ:kl_l1_inequ}
        D_{KL}(p \| q) = \sum_{i} p_i \log_2 \frac{p_i}{q_i} \ge \frac{1}{2 \ln 2} \| p - q \|_1^2 \, ,
    \end{equation}
    we have\footnotemark[1]
    \begin{align*}
        \| \mathcal{L} - \mathcal{L}^\prime \|_F^2 &= \sum_{i=1}^{n} d_i \cdot \| \tilde{A} - \tilde{A}^\prime \|_1^2 \\
        &\le 2\ln 2 \sum_{i=1}^{n} d_i \cdot D_{KL}(\tilde{A} \| \tilde{A}^\prime) \\
        &= 2 \cdot \sum_{i=1}^{n} d_i \cdot \sum_{j=1}^{n} \frac{A(i, j)}{d_i} \ln \frac{A(i, j)}{A^\prime(i, j)} \\
        &= 2 \cdot \sum_{i=1}^{n} \sum_{j=1}^{n} A(i, j) \ln \frac{A(i, j)}{A^\prime(i, j)} \\
        &= 2 \cdot L(D\mid M)
    \end{align*}
    \footnotetext[1]{The base of logarithm of Inequality \eqref{equ:kl_l1_inequ} is 2, thus the factor $\ln 2$ vanished since we change the base from 2 to $e$.}
    Together, we have
    \begin{equation}
        \sum_{i=1}^{n} ( \lambda(i) - \lambda^\prime(i) )^2 \le 2 \cdot L(D\mid M)
    \end{equation}
\end{proof}

\subsection{Proof of Theorem 3.2}
\begin{proof}
The key tool is Jensen's inequality. Suppose $f(x)$ is a convex function, then
\begin{equation}
    f(\frac{\sum_{i} \lambda_i x_i}{\sum_{i} \lambda_i}) \le \frac{\sum_{i} \lambda_i f(x_i)}{\sum_{i} \lambda_i} \, ,
\end{equation}
where $\lambda_i$ is positive weights.

In the following proof, the inequality is applied on $f(x) = -\ln x$.

{
\begin{equation}\label{equ:eq1}
    \begin{aligned}
        \Delta L_E(i, j) 
                &= 2 \left( D_i \ln \frac{D_i+D_j}{D_i} + D_j \ln \frac{D_i+D_j}{D_j} \right) \\
                &+ 2\sum_{l\ne i, j} \left( A_S(i, l)\ln \frac{A_S(i, l)}{A_S(i, l)+A_S(j, l)} + A_S(j, l)\ln \frac{A_S(j, l)}{A_S(i, l)+A_S(j, l)} \right) \\
                &+ A_S(i, i)\ln \frac{A_S(i, i)}{A_S(i, i)+A_S(j, j)+2A_S(i, j)} + A_S(j, j)\ln \frac{A_S(j, j)}{A_S(i, i)+A_S(j, j)+2A_S(i, j)} \\
                &+ 2 A_S(i, j) \ln \frac{A_S(i, j)}{A_S(i, i) + A_S(j, j) + 2A_S(i, j)}
    \end{aligned}
\end{equation}
}
Let's focused on the part related to $i$ in Equation \eqref{equ:eq1}, since $i$ and $j$ are symmetric.
\begin{align*}
        \Delta L_E(i, j)_{i} &= 2 D_i \ln \frac{D_i+D_j}{D_i} \\
        &+ 2 \sum_{l\neq i, j} \left( A_S(i, l) \ln \frac{A_S(i, l)}{A_S(i, l)+A_S(j, l)} \right) \\
        &+ A_S(i, i) \ln \frac{A_S(i, i)}{A_S(i, i)+A_S(j, j)+2A_S(i, j)} \\
        &+ A_S(i, j) \ln \frac{A_S(i, j)}{A_S(i, i)+A_S(j, j)+2A_S(i, j)} \\
        \intertext{Apply Jensen's inequality on the last two row}
        &\ge 2 D_i \ln \frac{D_i+D_j}{D_i} \\
        &+ 2 \sum_{l\neq i, j} \left( A_S(i, l) \ln \frac{A_S(i, l)}{A_S(i, l)+A_S(j, l)} \right) \\
        &+ 2(A_S(i, i)+A_S(i, j)) \cdot \left(-\ln \frac{A_S(i,i)+A_S(j,j)+2A_S(i,j)}{A_S(i, i)+A_S(i, j)} \right) \\
        &= 2D_i \bigg[ \sum_{l\neq i,j} \frac{A_S(i, l)}{D_i} \cdot \left( -\ln \frac{A_S(i, l)+A_S(j,l)}{A_S(i,l)} \frac{D_i}{D_i+D_j} \right) \bigg. \\
        &+ \bigg. \frac{A_S(i,i)+A_S(i,j)}{D_i} \cdot \left( -\ln \frac{A_S(i,i)+A_S(j,j)+2A_S(i,j)}{A_S(i,i)+A_S(i,j)} \frac{D_i}{D_i+D_j} \right) \bigg] \\
        \intertext{Apply Jensen's inequality again}
        &\ge 2D_i \cdot \left( -\ln \left[ \sum_{l\neq i, j} \frac{A_S(i,l)+A_S(j,l)}{D_i+D_j} + \frac{A_S(i,i)+A_S(j,j)+2A_S(i,j)}{D_i+D_j} \right] \right) \\
        &= 2 D_i \cdot (-\ln 1) \\
        &= 0
\end{align*}
Similarly, $\Delta L_E(i, j)_{j} \ge 0$. Together, $\Delta L_E(i, j) \ge 0$.
\end{proof}

\bibliographystyle{siamplain}
\bibliography{ref_appendix}